\documentclass[apj]{emulateapj}
\usepackage{float,epsfig}
\usepackage{pstricks}

\newcommand{\mincir}{\raise
  -2.truept\hbox{\rlap{\hbox{$\sim$}}\raise5.truept \hbox{$<$}\ }}
\newcommand{\magcir}{\raise
  -2.truept\hbox{\rlap{\hbox{$\sim$}}\raise5.truept \hbox{$>$}\ }}
\begin{document}

\submitted{The Astrophysical Journal, 646:143--160, 2006 July 20}
\title{The Two-dimensional XMM-Newton Group Survey: $z<0.012$ groups.}
\author{Alexis Finoguenov\altaffilmark{1,2}, David S. Davis\altaffilmark{2,3}, Marc
Zimer\altaffilmark{1}, John S. Mulchaey\altaffilmark{4}}

\altaffiltext{1}{Max-Planck-Institut f\"ur Extraterrestrische Physik,
             Giessenbachstra\ss e, 85748 Garching, Germany}
\altaffiltext{2}{University of Maryland, Baltimore County, 1000
  Hilltop Circle,  Baltimore, MD 21250, USA}

\altaffiltext{3}{Laboratory for High Energy Astrophysics, NASA Goddard Space
  Flight Center,  Code 661.0, Greenbelt, MD 20771; ddavis@milkyway.gsfc.nasa.gov}
\altaffiltext{4}{Observatories of the Carnegie Institution of Washington, 813
Santa Barbara Street, Pasadena, CA 91101; mulchaey@ociw.edu}  

\begin{abstract}

We present the results of the 2-dimensional XMM-Newton Group Survey (2dXGS),
an archival study of nearby galaxy groups. In this paper we consider eleven
nearby systems ($z<0.012$) in Mulchaey et al. (2003), which span a broad
range in X-ray luminosity from $10^{40}$ to $10^{43}$ ergs/s. We measure the
iron abundance and temperature distribution in these systems and derive
pressure and entropy maps.  We find statistically significant evidence for
structure in the entropy and pressure of the gas component of seven groups
on the 10--20\% level. The XMM-Newton data for the three groups with best
statistics also suggest patchy metalicity distributions within the central
20--50 kpc of the brightest group galaxy, probed with 2--10 kpc
resolution. This provides insights into the processes associated with
thermalization of the stellar mass loss. Analysis of the global properties
of the groups reveals a subclass of X-ray faint groups, which are
characterized by both higher entropy and lower pressure.  We suggest that
the merger history of the central elliptical is responsible for both the
source and the observed thermodynamical properties of the hot gas of the
X-ray faint groups.

\end{abstract}


\section{Introduction}

Groups of galaxies constitute the most common galaxy association, containing
as much as 50--70\% of all galaxies in the nearby universe (Geller \& Huchra
1983) and provide a link between the massive virialized systems such as
clusters of galaxies and the field.  For example, Finoguenov et al. (2003b)
found that chemical enrichment in groups is very close to the prediction of
the Salpeter Initial Mass Function, known to characterize the star-formation
in the field. It was also suggested by Kodama et al. (2001) that properties
of galaxies in clusters are largely defined by their previous group
environment. Thus, studies of groups open prospects to understanding galaxy
formation, presently one of the most challenging astrophysical issues.

Although the systems we study are defined by galaxy concentration, the
presence of X-ray emission introduces further refinement in the definition
and indicates that some of these systems are also {\it virialized} objects
(Ostriker et al. 1995). Groups with a high fraction of early-type galaxies
are more likely to exhibit detectable X-ray emission (Mulchaey et
al. 1996). However, at low X-ray luminosities ($\sim10^{41}$ ergs s$^{-1}$)
the emission from individual galaxies becomes a non-negligible contribution
to the overall diffuse emission. Early-type galaxies show a wide range of
X-ray properties, from the ``normal'' emission to those early-type galaxies
that have rather compact X-ray emission (Forman et al. 1985; Matsushita
2001; Finoguenov \& Miniati 2004), while others have been suggested to lack
a massive dark matter halo (Romanowsky et al. 2003, but see Dekel et
al. 2005).  Group gas, consisting of bona fide collapsed halos, obey
well-defined X-ray scaling relations, and one can use these relationships to
shed light on the large-scale dark matter distributions in these systems.

\begin{figure*}
\includegraphics[width=8.cm]{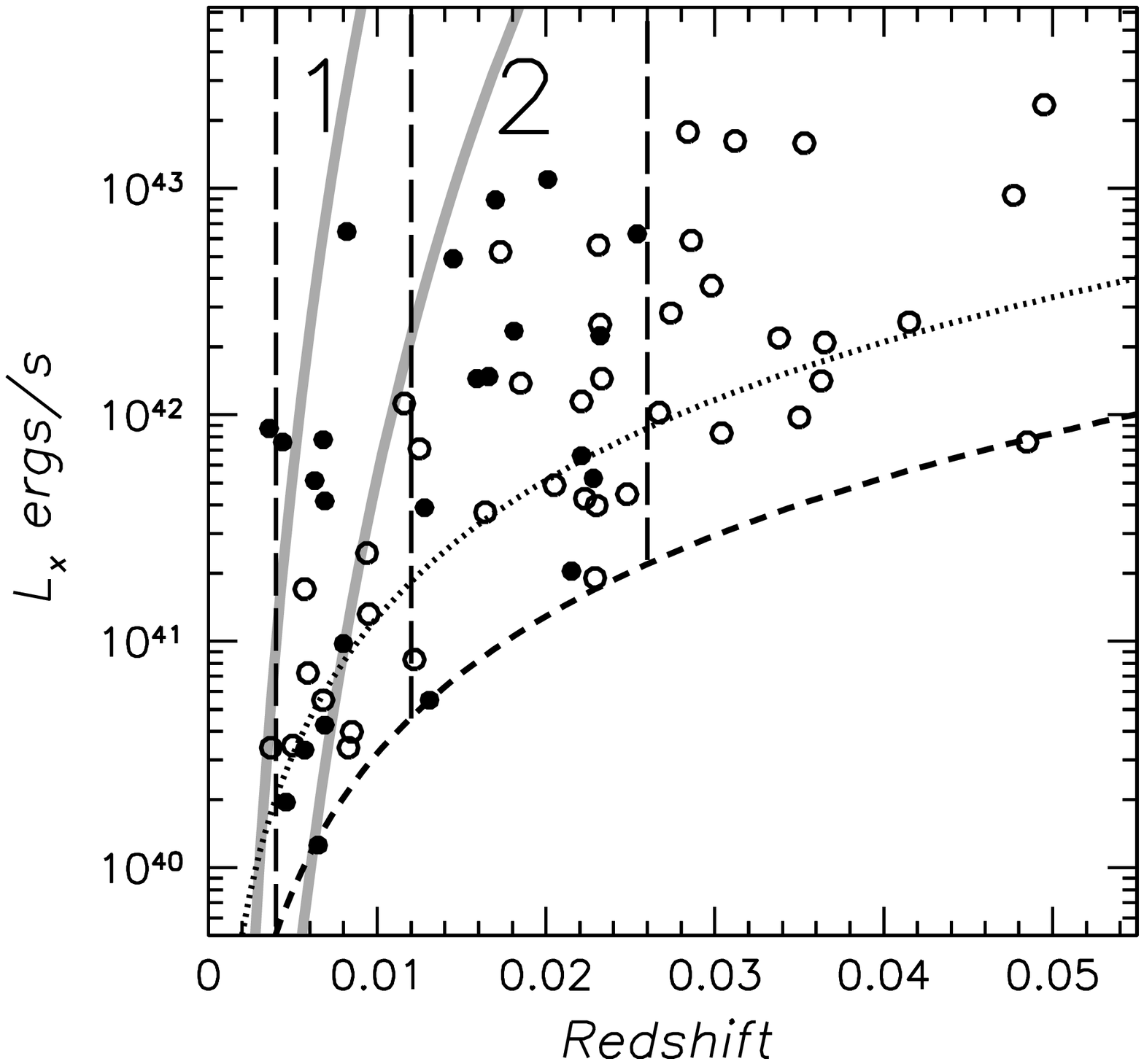}\includegraphics[width=8.cm]{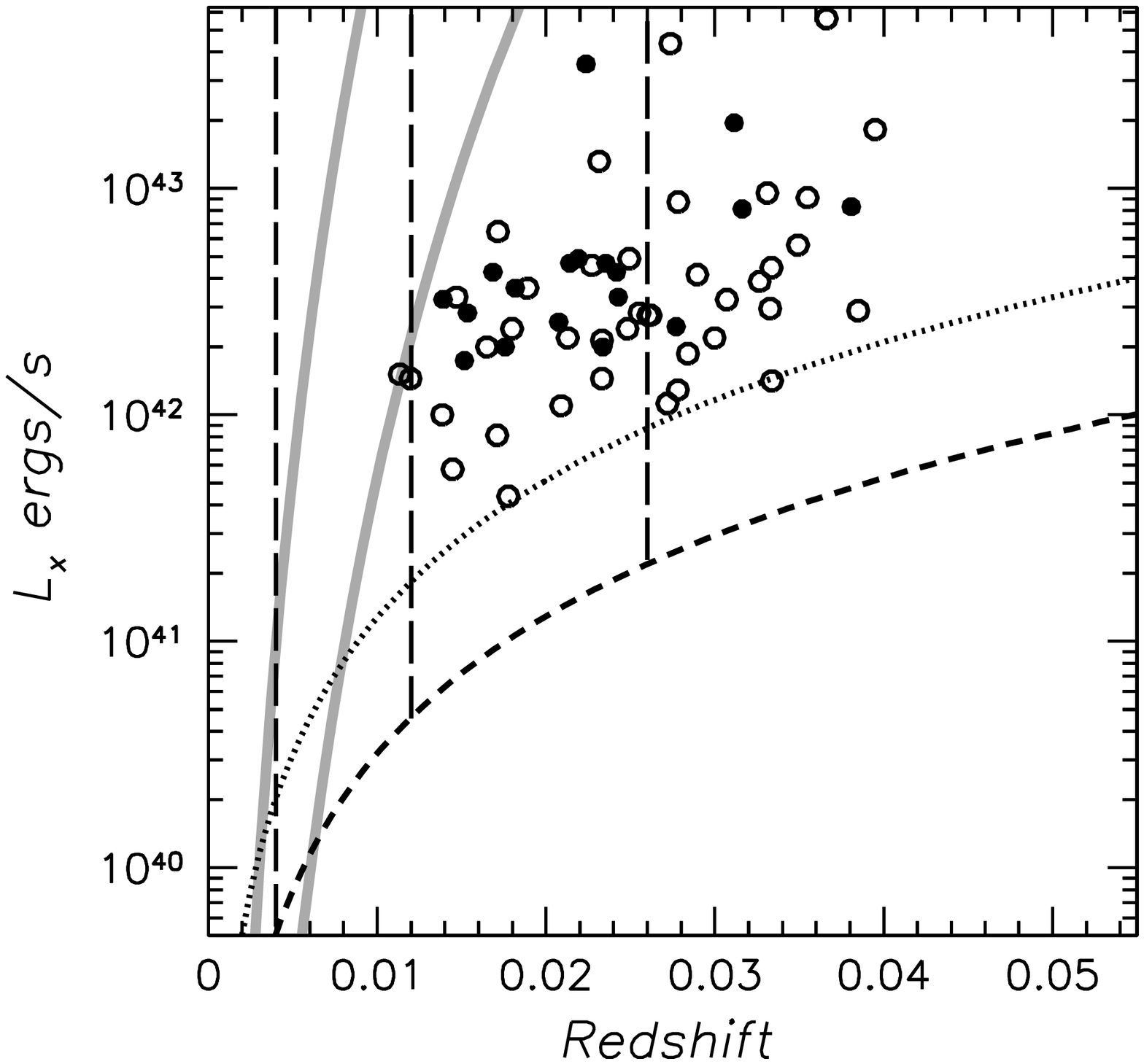}
\figcaption{{\it Left panel}. Observed characteristics of the 2DXGS. The
circles correspond to groups listed in Mulchaey et al. (2003). Filled
circles denote objects observed by XMM-Newton.  The change in the luminosity
limit corresponding to the same X-ray flux in correspondence to Mulchaey and
Mahdavi surveys is shown with short-dashed and dotted lines,
respectively. The vertical long-dashed lines define the two subsamples in
2DXGS and are labeled 1 and 2. The analysis of the first subsample is
reported here. The XMM field-of-view will enclose 1/4 of the virial radius
for the groups to the right of the low redshift gray line and 1/2 of the
virial radius for the higher redshift gray line. The quoted luminosities are
bolometric assuming $h_{100}$ (Mulchaey et al. 2003) and are measured out to
the radius of X-ray detection.  {\it Right panel}. The same as the left
panel, but for the Mahdavi et al. (2000) group sample. The luminosities are
given for 0.1--2.4 keV band, $h_{100}$ and aperture of $500$ kpc (Mahdavi et
al. 2000).  For systems in common with the Mulchaey et al. (2003) sample,
these luminosities are lower by 0.1--0.2 dex, which is accounted for by the
bolometric correction.
\label{f:survey}}
\end{figure*}

Comparative studies of the scaling relations in clusters of galaxies reveal
strong deviations of the observed relations from predictions based on
self-similar collapse (e.g. Ponman et. al. 1996; Finoguenov et
al. 2001). These deviations are thought to be best characterized by the
injection of energy (preheating) into the gas before clusters collapse
(Kaiser 1991; Evrard \& Henry 1991).  This entropy `floor' causes the simple
self-similar scaling laws (Kaiser 1986) to predict a shallower $L_X - T$
relationship. However, analysis of a large compilation of entropy profiles
on groups and clusters of galaxies is inconsistent with this picture,
requiring at $r_{500}$ much larger entropy levels than were thought before
(Finoguenov et al. 2002) and modifying the concept of the entropy floor to
the entropy ramp at $0.1r_{200}$ (Ponman et al. 2003). Reproduction of these
results both analytically and numerically, strongly supports the scenario of
Dos Santos \& Dore (2002), where an initial adiabatic state of the infalling
gas is further modified by the accretion shock (Ponman et al. 2003; Voit et
al. 2003; Voit \& Ponman 2003; Borgani et al. 2005).  As supporting
evidence for the latter, Ponman et al. (2003) noticed a self-similarity in
the entropy profiles, once scaled to $T^{0.65}$. Some XMM-Newton
observations are consistent with this result (Pratt \& Arnaud 2003;
Finoguenov et al. 2004b; Finoguenov et al. 2005a). A major change introduced
by these studies is that groups of galaxies can again be viewed as
scaled-down versions of clusters, but with the scaling itself modified. Further
evidence for the departure of groups from the trends seen in clusters has
been recently refuted by Osmond \& Ponman (2004).

In addition to preheating, internal heating by supernovae and AGN may also
impact the observed scaling relations. Recent results from Chandra
demonstrate that AGN heating is likely very important in both the centers of
clusters (McNamara et al. 2001) and early-type galaxies (Finoguenov \& Jones
2001). However, very little work has been done to explore the role of AGN
heating on the intragroup medium.

To better understand the physics of the intragroup medium, we have embarked
on an XMM-Newton archival study of nearby galaxy groups. The large field of
view of XMM-Newton, combined with its good resolution and large collecting
area make it an ideal instrument to study the hot gas in nearby groups.  We
also perform a study of how structure in the observed systems may affect
their placement on the scaling relations for entropy. We adopt a Hubble
constant of 70 km s$^{-1}$ Mpc$^{-1}$, $\Omega_M=1-\Omega_\Lambda=0.3$, and
present redshifts corrected for the local motion.

\section{Sample Selection}

At the moment there is no large purely X-ray selected sample of groups
of galaxies. For example, only a few objects, Fornax cluster, MKW4,
NGC~4636, NGC~1550 and NGC~5044 are present in the HIFLUGS, a complete
all-sky sample of brightest groups and clusters of galaxies (Reiprich
\& B\"ohringer 2002).  Most present-day samples of groups are based on
the X-ray follow-up of optical surveys (e.g. Mahdavi et al. 2000;
Mulchaey et al. 2003).

For this study we have selected the groups in Mulchaey et al. (2003) with
publicly available XMM-Newton (Jansen et al. 2001) observations.  Most of
the groups in the Mulchaey et al. (2003) sample were found by
cross-correlating the ROSAT observation log with the positions of
optically-selected groups in the catalogs of Hickson (1982), Huchra \&
Geller (1982), Geller \& Huchra (1983), Maia et al. (1989), Nolthenius
(1993) and Garcia (1993). In addition to the optically-selected groups, the
Mulchaey et al. (2003) sample includes a small number of groups that were
discovered in the ROSAT All-Sky Survey (RASS). Their final group list
contains 109 systems.

The properties of groups from the Mulchaey et al. (2003) sample are shown in
Fig.\ref{f:survey}. The flux limit corresponds to $2.6\times10^{-13}$ ergs
s$^{-1}$ cm$^{-2}$ at a 3$\sigma$ detection level, which is a factor of 4
lower than the flux limit in the Mahdavi et al. (2000) sample, a factor of
10 lower than in the REFLEX/NORAS surveys and a factor of 100 lower compared
to HIFLUGS (these latter surveys are all based on the RASS data). The
distribution of systems in Fig.\ref{f:survey} is quite homogeneous, as are
the targets observed by XMM-Newton (which covers 50\% of groups at redshifts
lower than 0.025; independent of the volume and luminosity cut).  By
construction, the Mulchaey et al. (2003) group list covers a fraction of a
percent of the sky for serendipitous systems, yet it includes a larger
fraction of bright objects, that were selected as primary targets for ROSAT
pointings, e.g. 3 of the 4 HIFLUGS groups are included.

As we are interested in studying the spatial variations in the intragroup
medium, we limit our analysis here to nearby groups. However, for the very
nearest groups in the Mulchaey et al. (2003) sample, only the very centers
of the group are covered in a single XMM-Newton pointing. Thus, we restrict
our sample to the redshift bin 0.004--0.012. Results for the higher redshift
bin 0.012--0.025 are presented in Finoguenov et al. (2005b). In addition, eight
groups from the Mahdavi sample are analyzed in Mahdavi et al. (2005). Two of
the XMM-Newton targets in our redshift bin had no diffuse emission detection
with ROSAT (NGC~4168 and NGC~7582), but we include them here for
completeness.

\begin{table}
\begin{center}
\renewcommand{\arraystretch}{1.}\renewcommand{\tabcolsep}{0.05cm}
\caption{\footnotesize
\centerline{Basic properties of the sample.}
\label{t:b}}

\begin{tabular}{llrllllll}
\hline
\hline
Name & $\sigma_{gr}$ &  & $r_e$ & log         & $\sigma_{c}$ &
\multicolumn{2}{l}{log $L_x$ \  \  \  \  z$_{\rm BCG}$} & $r_{500}$\\
     &  km/s & $N_{gr}$   & kpc   &$L_{B, \odot}$& km/s        & \multicolumn{2}{l}{ergs/s  \  \  \  \   $10^{-4}$}& kpc\\
\hline
{\footnotesize NGC~2300} &$278^{+35}_{-31}$    & 16 &     & 10.41 &254$\pm$23& $41.93\pm0.04$ & 65&339\\
{\footnotesize NGC~3665} &$45^{+48}_{-44}$     & 6  &     & 10.70 &186$\pm$11& $40.94\pm0.37$ & 69&222\\ 
{\footnotesize IC~1459}  &$256^{+44}_{-38}$    & 11 & 4.4 & 10.37 &308$\pm$7 & $40.83\pm0.36$ & 56&280\\
{\footnotesize NGC~3923} &$191^{+126}_{-115}$  & 9  & 13. & 10.52 &275$\pm$23& $40.60\pm0.37$ & 58&280\\
{\footnotesize NGC~4168} &$259^{+56}_{-52}$    & 10 & 6.9 & 10.40 &182$\pm$11& $<40.87$       & 76&258\\
{\footnotesize NGC~4261} &$429^{+54}_{-50}$    & 43 & 5.5 & 10.70 &294$\pm$16& $42.20\pm0.07$ & 75&451\\
{\footnotesize NGC~4636} &  n/a                & 1  & 7.1 & 10.51 &191$\pm$12& $42.19\pm0.20$ & 31&331\\
{\footnotesize NGC~5044} &$357^{+48}_{-42}$    & 15 & 19. & 10.70 &238$\pm$2 & $43.12\pm0.02$ & 90&430\\
{\footnotesize NGC~5322} &$198^{+124}_{-117}$  & 7  & 5.2 & 10.67 &224$\pm$5 & $40.41\pm0.38$ & 59&283\\
{\footnotesize NGC~5846} &$368^{+51}_{-46}$    & 20 & 9.2 & 10.66 &278$\pm$23& $42.02\pm0.17$ & 57&309\\
{\footnotesize NGC~7582} &$38^{+37}_{-36}$     &  8 &     & &   &              $<41.05$       & 54&   \\
\hline
\end{tabular}
\end{center}
\end{table}

The basic properties of our sample groups are listed in
Tab.\ref{t:b}. Col.(1) gives the name of the group, (2) velocity dispersion
of the group, (3) number of galaxies identified with the group, (4--6)
properties of the central galaxy: effective radius, blue luminosity and
central velocity dispersion, (7) X-ray luminosity scaled to $h_{70}^{-2}$ as
tabulated in Mulchaey et al. (2003), (8) redshift, (9) $r_{500}$ in kpc,
{ calculated as $r_{500}=0.391 {\rm Mpc} \times (kT/{\rm
keV})^{0.63}h_{70}^{-1}$ using the M--T relation (Pacaud, F. 2005, private
communication) re-derived from Finoguenov et al. (2001) using an
orthogonal regression and correcting the masses to $h_{70}$ and a LCDM
cosmology.} The optical data are collected from Bender et al. (1992),
Koprolin \& Zeilinger (2000) and Davies et al. (1987). NGC~3665 and NGC~2300
are classified as S0 and NGC~7582 is classified as a spiral (de Vaucouleurs
et al. 1991; RC3 hereafter).  After a careful investigation, we could not
confirm the members of the NGC~4636 group (see section 5.7).

\section{Data Analysis}

Tab.\ref{t:ol} lists details of the observations. Column (1) lists the name
of the group, (2) is the assigned XMM archival name, (3) is the net Epic-pn
exposure after removal of flaring episodes, (4) lists the pn filter used,
which determines the choice of instrumental response as well as background
estimates, (5) the XMM-Newton revolution number, which is useful to assess
any secular evolution of the instrumental background, and (6) the pn frame
time, which is used to determine the fraction of out-of-time events.

\begin{table}
\begin{center}
\renewcommand{\arraystretch}{1.1}\renewcommand{\tabcolsep}{0.12cm}
\caption{\footnotesize
\centerline{Log of the XMM Epic-pn observations of groups.}
\label{t:ol}}

\begin{tabular}{rccccr}
\hline
\hline
 Name &  Obs.       &             net    &  pn     & XMM   &frame\\
      &   ID        &             exp.    &  Filter & Orbit &time\\
      &             &             ksec    &         &       &ms\\
\hline
 NGC~2300     & 0022340201  &   42.3 & Thin   & 232 &  73\\
 NGC~3665     & 0052140201  &   19.5 & Medium & 363 & 199\\
  IC~1459     & 0135980201  &   26.3 & Medium & 438 &  73\\
 NGC~3923     & 0027340101  &   30.3 & Thin   & 379 &  73\\
 NGC~4168     & 0112550501  &   15.8 & Medium & 364 & 199\\
 NGC~4261     & 0056340101  &   20.3 & Medium & 370 &  73\\
 NGC~4636     & 0111190701  &   55.7 & Medium & 197 &  73\\
 NGC~5044     & 0037950101  &   10.5 & Medium & 201 &  73\\
 NGC~5322     & 0071340501  &   13.6 & Thin   & 374 &  73\\
 NGC~5846     & 0021540101  &   28.6 & Thin   & 207 &  73 \\
 NGC~7582     & 0112310201  &   17.8 & Medium & 267 &  73 \\
\hline
\end{tabular}
\end{center}
\end{table}

The initial steps of the data reduction are similar to that described
in Zhang et al. (2004) and Finoguenov et al. (2003a). One important
aspect of the data preparation is to remove flares which can
significantly enhance the detector background and severely limit the
detection of low surface brightness features. Thus, for the group
analysis, using flare free periods is critical.  To identify time intervals
with flares we look at energies above 10 keV where the telescope efficiency 
is quite low and the particle background
dominates the source counts. We use the 10--15 keV energy band
(binned in 100 s intervals) to monitor the particle background and to
excise periods of high particle flux. In this screening process we use
the events with FLAG=0 and PATTERN$<5 (13)$ for the pn (MOS) data.  We
reject time intervals affected by flares by excising periods where the
detector count rate exceeds 2$\sigma$ above the mean quiescent rate.

Since most of the observations analyzed here were performed using a short
integration frame time for pn, it is important to remove the out-of-time
events (OOTE) for accurate imaging and spectral analysis. We used the
standard product of the {\it epchain} SAS (XMMSAS 6.1) task to produce the
simulated OOTE file for all the observations and scaled it by the fraction of
the OOTE expected for the frame exposure time, as specified in
Tab.\ref{t:ol}. 

{ For the broad-band analysis we have used both pn and MOS data, while in
the spectral analysis we only use pn data in order to take advantage of our
expertise on the background subtraction for pn.}  In addition, the pn
spectra have the highest statistics for the groups. We use response
matrices released under XMMSAS 6.1. For further details of XMM-Newton
processing we refer the reader to 
{\it http://wave.xray.mpe.mpg.de/xmm/cookbook/general}.

Analysis of the groups consists of two main steps: estimating the
temperature structure of groups and verifying the structure using spectral
analysis. Only the results of the spectral analysis are used in tabulations
and comparison studies. The initial maps based on the hardness ratios are
only used to locate the substructure. Quantitative analysis of the
substructure, possible only for a subsample of systems is based on the
spectral analysis. The two dimensional information, related to 
the analysis reported in this paper, is released under 
http://www.mpe.mpg.de/2dXGS/ homepage.

The first part of the analysis consists of producing temperature estimates
for each of the systems. We use the surface brightness in conjunction with
the hardness ratio maps to locate regions of similar intensity and
color. The input data are based on images corrected for the instrumental
effects discussed above and are background subtracted. The surface
brightness is a wavelet-reconstructed (Vikhlinin et al. 1998) image in the
0.5--2 keV band and the hardness of the emission is a ratio of the
wavelet-reconstructed images in the 0.5--1 and 1--2 keV bands. An advantage
of using wavelets is the ability to remove additional background by spatial
filtering as well as control over the statistical significance of the
detected structure. We apply a four sigma detection threshold, followed out
to a 90\% confidence limit, 'a trous' method of wavelet image reconstruction
with scales from $8^{\prime\prime}$ to $4^\prime$. Complications arise since
the wavelet algorithm splits the image into discrete scales resulting in
small scale discontinuities in the reconstructed image, which we overcome by
applying additional smoothing before producing the hardness ratio map.  The
point sources are not removed in the imaging analysis.

We also construct pseudo entropy and pressure maps using the image
and the hardness ratio map. Although these maps suffer from a number of
degeneracies, with metalicity-density being the strongest since a
significant fraction of the group X-ray emission is due to lines, they
indicate the regions of primary interest for the detailed spectroscopic
analysis which allows most of the degeneracies to be resolved. For a system
in hydrostatic equilibrium, we expect the entropy to monotonically increase
with the increasing radius (e.g. Metzler \& Evrard 1994), while the pressure
should decrease with radius.

The spectroscopic part of the analysis uses a mask file. This mask is
created using the results of the hardness ratio and surface brightness
analysis described above. The first application of this technique is
presented in Finoguenov et al. (2004a). In our analysis we wish to select
contiguous regions with similar spectral properties. This allows us to
combine regions so that counting statistics are not the limiting factor in
our derivation of the group properties. To select regions we use the wavelet
based hardness ratio maps to identify regions with similar X-ray colors and
similar intensity level. To generate the mask file for use in further
spectral analysis we use hardness ratios that correspond to temperatures in
the ranges 0.48-0.56-0.60-0.64-0.72-0.8-0.9-1.0-1.1-1.2~keV and have
intensities that are the same within a factor of two. We then examine each
of the isolated regions with approximately equal color and intensity
imposing the additional criterion that the regions should be larger than the
PSF (15\arcsec ) width and contain more than 300 counts in the raw pn image.

To demonstrate the performance of the technique, for NGC~4636 we compare the
temperatures obtained in the spectral analysis with the temperatures
estimated from the hardness ratio map within the central $5^\prime$ radius
(Fig.\ref{f:tcomp}). The mask file is used to extract the spectra and is
used to estimated the mean and dispersion of the expected temperature from
the hardness ratio map. The values from the spectral analysis are taken as
the best-fit value and we use the corresponding 68\% confidence interval. We
find that the estimated and measured temperatures agree to within 20\%.  The
spectral and hardness ratio temperatures show a systematic offset below
about 0.85 keV with the hardness ratio temperatures being $\sim$ 20\% cooler
than the fitted temperatures.  The scatter about the best fit line is less
than $\sim$10\%, which matches our errors in determining the temperature.
This shows that the hardness ratio in the 0.5--1 to 1--2 keV bands reflect
the temperature of the gas.  The cause of systematic deviations from unity
is likely due to the higher absorption column assumed in determining the
hardness ratio to temperature conversion.  The systematic offset between the
hardness ratio temperature and the spectral temperature is such that it
could result from uncertainties in the Fe abundance. The tabulated results
are from the spectral fits and the column density and abundance are fitted.

\includegraphics[width=8.5cm]{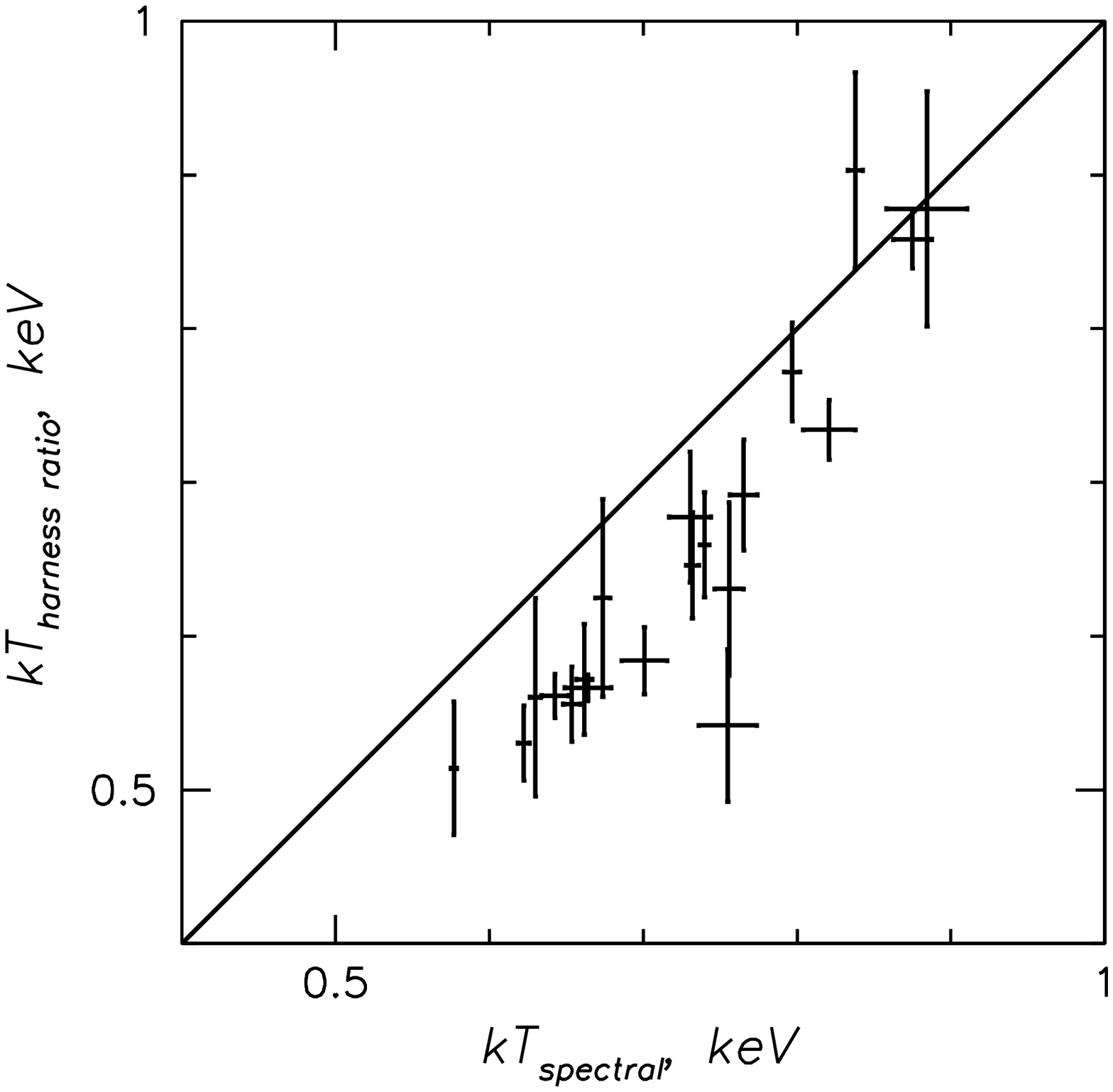}

\figcaption{Comparison between the temperature derived from the
spectral analysis and the temperature estimated from the hardness
ratio in NGC~4636. 
\label{f:tcomp}}

The spectral analysis was performed using the single-temperature APEC plasma
code, fitting the abundance of O--Ni as one group with their relative
abundances at the photospheric solar abundance ratios of Anders \& Grevesse
(1989). Absorption was fixed at the Galactic value, reported in
Tab.\ref{t:prop}.

In the spectral analysis, special attention was paid to the issue of the
background estimation. We employ a double-subtraction technique, following
Zhang et al. (2004). We used the 12--16$^\prime$ radius and 2--12 keV band
to estimate the quality of the background subtraction using the blank
fields.  Variations in the background are found on the 20\% level, in
agreement with other studies (Freyberg et al. 2004).  In our analysis of
groups, we fix the shape of the background component and add a 0.2 keV
thermal (APEC with solar element abundance) component to account for a
possible variation in the Galactic foreground, allowing the normalizations
of both components to be fit. Better estimation of the background allows one
to carry out the spectral analysis throughout the detector if emission is
found or set strict upper limits otherwise. Typically, an improvement of
$\chi^2$ is achieved for luminous parts of the group.

The metal abundance for the plasma model was explored using three
approaches: element abundance fixed at 0.3 solar value, variable element
abundance, variable abundance plus a 1.5 slope power law component to
account for the contribution of low mass X-ray binaries (LMXBs). We use
0.4--5 keV band for fitting, and group the channels to achieve 30 counts per
bin to use the $\chi^2$ statistic. In two cases (IC~1459, NGC~4261) an
additional component was added to fit the contribution of the AGN in the
center of the group. All other identified point sources were excised from
the spectral extraction. Some regions, with too few counts for spectral
analysis and that could not be combined with other regions on the basis of
color or intensity, are not analyzed and appear as excised sources. The
results obtained with the fixed element abundance are only used to evaluate
the temperature structure when statistics are poor. Results from these fits
are not used to derive any of the quantities given in the tables.

To estimate the pressure and entropy in each region, we need to estimate the
length of the column for each selected two-dimensional region on the sky. We
assume a geometry in which every region is a part of a spherical shell that
is centered on the system core and has its inner and outer radii passing
through the nearest and furthest points of the selected region,
respectively. This spherical shell is further intersected by a cylinder,
that is directed towards the observer and in the observer plane has the
cross-section of the selected region. For the concentric regions our
geometry corresponds to a usual "onion peeling'' technique (e.g. Finoguenov
\& Ponman 1999).  We assume a constant gas distribution with the integrated
emissivity in the region matched to the measured emissivity in the spectral
analysis. With these approximations the longest length through each volume
is $L=2 \sqrt{r_{max}^2 - r_{min}^2}$. The volume of the region is then $2/3
S \times L$, where S is the area of the region. { For reasonable assumptions
of the geometry, the systematic uncertainty introduced by the volume
estimate is 5\%.} Further details can be found in Henry et al. (2004) and
Mahdavi et al. (2005). We also analyzed regions using concentric annuli.
The data for the annuli are extracted after { excising point sources as well
as the diffuse background sources identified in the text (NGC4168) and the
substructure exceeding 50\% (important for NGC 2300, NGC 5322, NGC4636) of
the expected value}. The annular analysis is used mainly to sample the
outermost regions as well as to provide a comparison with the standard
technique of volume estimate.

\section{Discussion \& Results}

Early examples of two-dimensional studies of groups were carried out using
ROSAT PSPC observations, which provided a reasonable sensitivity to both the
group temperature, as well as group luminosity. Some of these include M86
(e.g. Finoguenov \& Jones 2000), NGC~2300 (Mulchaey et al. 1993; Davis et
al. 1996), NGC~4261 (Davis et al. 1996), NGC~4636 (Trinchieri et al. 1994),
NGC~5044 (David et al. 1994), NGC~5846 (Finoguenov et al. 1999). For a more
complete review we refer the reader to Mulchaey (2000). The XMM data cover a
different spatial scale, different energy range, and the PSF is much
different than that used in the earlier ROSAT analysis. Thus, a comparison
of our results to those in the literature helps to determine how the derived
X-ray characteristics of the groups are affected by the characteristics of
different instruments.

\subsection {Global Properties}

\begin{figure*}
\includegraphics[width=8.5cm]{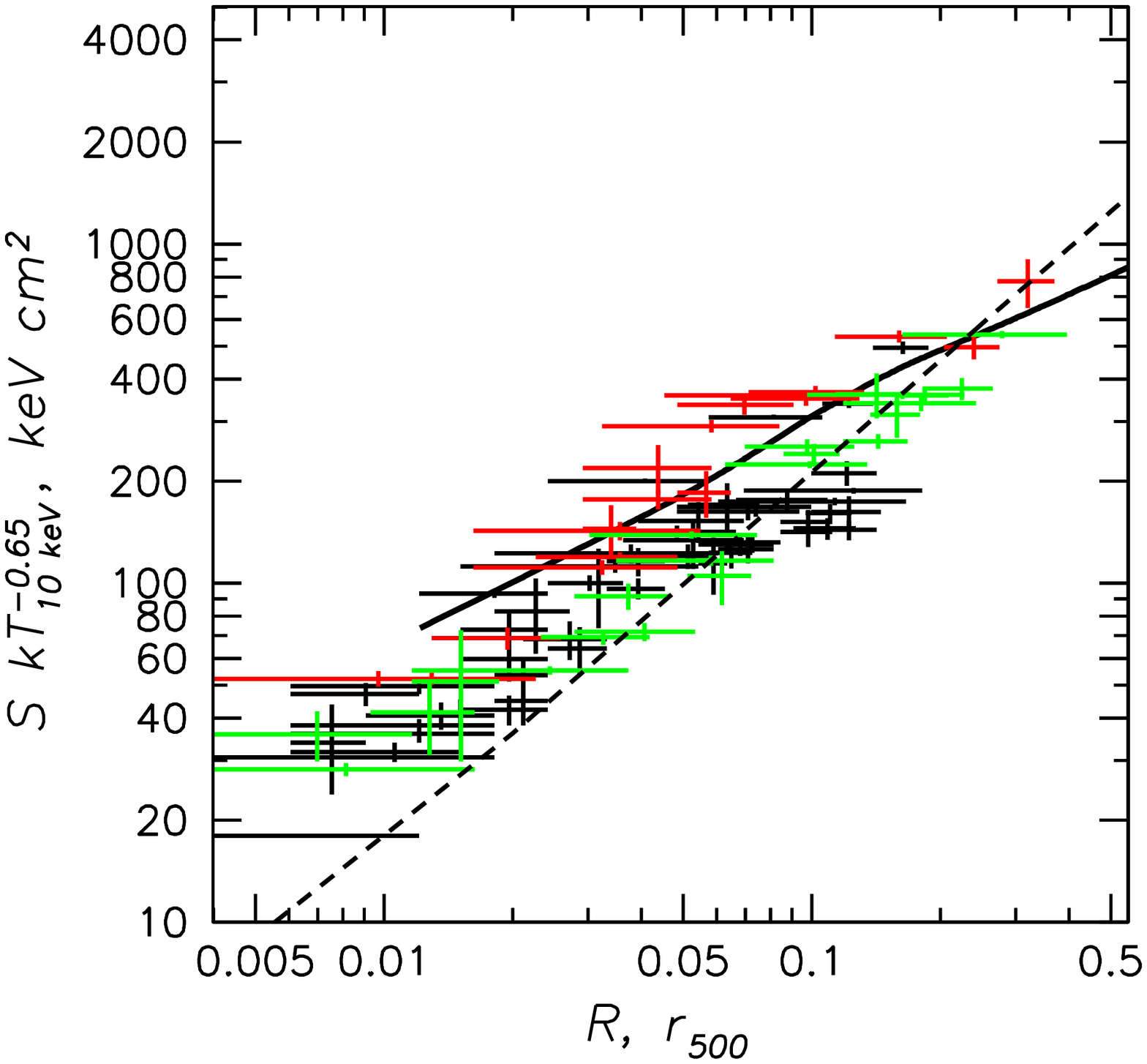}\hfill
\includegraphics[width=8.5cm]{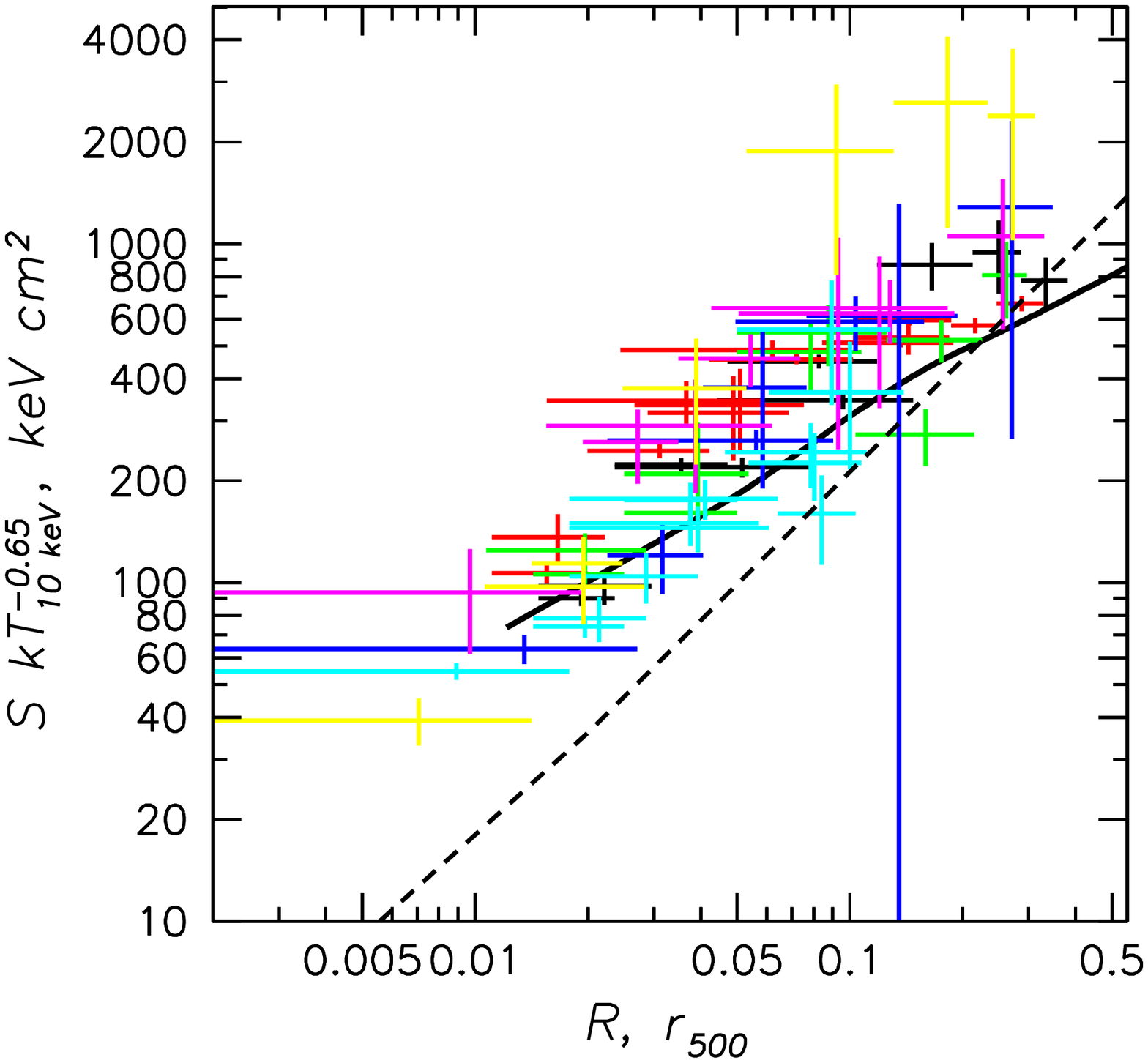}

\includegraphics[width=8.5cm]{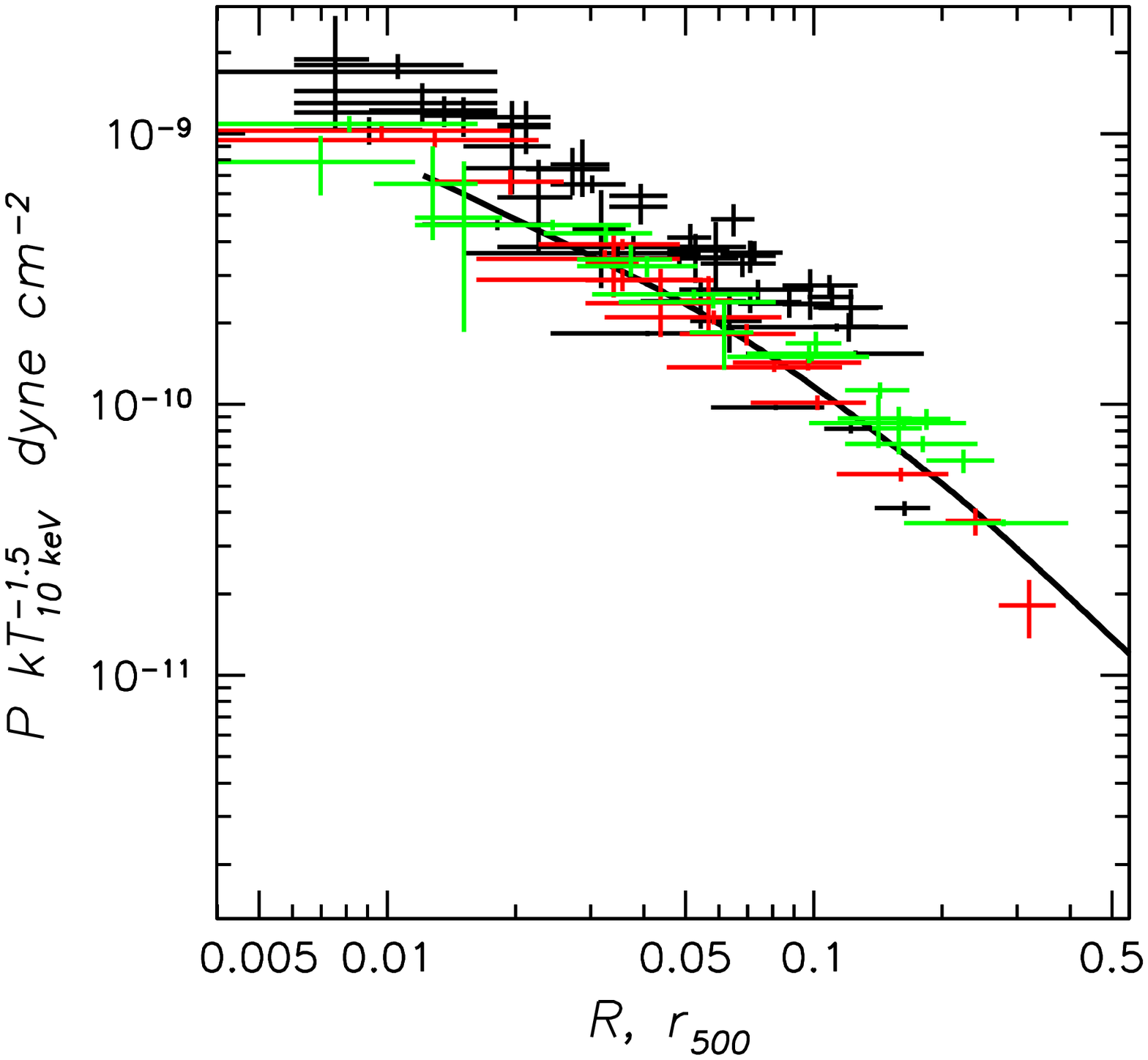}\hfill
\includegraphics[width=8.5cm]{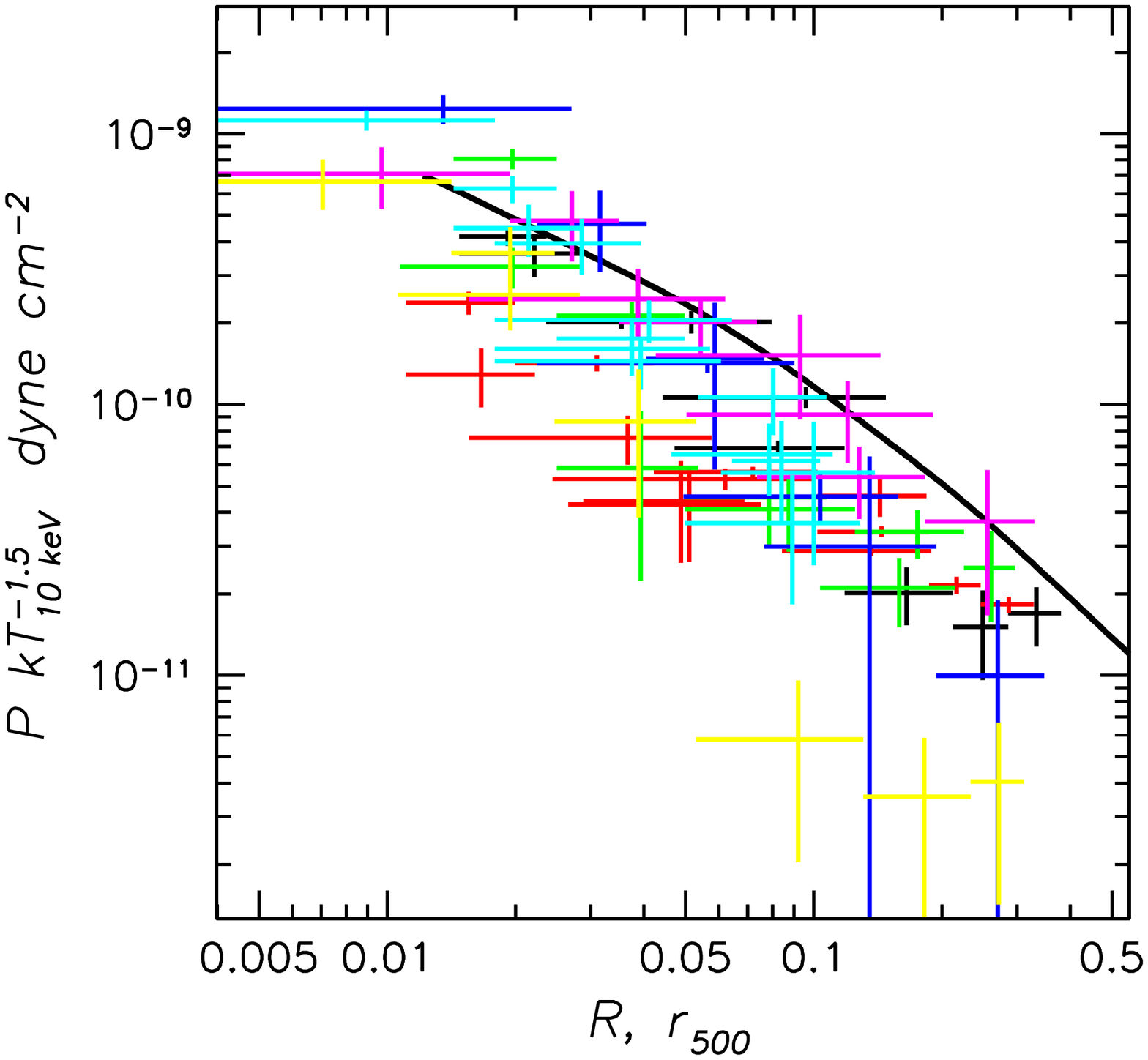}

\figcaption{Scaled entropy (top) and pressure (bottom) profiles for the
sample, { obtained from spectral analysis}. Left panel: NGC~4636 (black),
NGC~5846 (red), NGC~5044 (green). Right panel: NGC~2300 (black), NGC~4261
(red), IC~1459 (green), NGC~3665 (blue), NGC~4168 (magenta), NGC~3923
(cyan), NGC~5322 (yellow). The scaling for entropy and pressure is done in
accordance with the results of Ponman et al. (2003).  The dashed line on
both entropy panels shows the $r^{1.1}$ law, expected in theory. The solid
lines on each panel represents the typical profiles of X-ray luminous groups
from Finoguenov et al. (2005b) from two other samples shown in
Fig.\ref{f:survey}.
\label{cf:ent}}

\end{figure*}

Before proceeding with describing the structure of the systems in our study,
in this section we provide an overview on the global properties of the
groups. This is done in several different ways. A traditional comparison is
presented in Tab.\ref{t:prop}, where we estimate the temperature, luminosity
and element abundance from a single spectrum, extracted using a circular
region with a radius of 30 kpc. In Tab.\ref{t:prop}--\ref{t:prop2} we list
the properties of the gas within various fractions of $r_{500}$. In
Fig.\ref{cf:ent} we present the entropy and pressure profiles derived using
the extracted spectra and compare them with the known scaling relations. We
show both the values derived using annuli, where we excise the zones
associated with galaxies other then the central galaxies and deviations
exceeding 30\% (e.g. a background galaxy in NGC~4168; a spiral galaxy in
NGC~2300; a tail in NGC~4636), and the values determined using the maps,
plotting the corresponding values vs their distance from the group's center.
The latter comparison is the most sophisticated and provides considerably
more insight into the state of the gas than the more traditional approaches.
We have divided the sample into two groups for presentation (
Fig.\ref{cf:ent}). While we provide a careful discussion of the trends
below, one can already see some drastic differences in both entropy and
pressure between
the groups NGC~5044, NGC~5846 and NGC~4636 and the rest of the sample. { For
comparison, In Fig.\ref{cf:ent} we also show the typical entropy and
pressure profiles as observed by XMM-Newton in galaxy groups of bolometric
luminosity well in excess of $10^{42}$ erg/s (Finoguenov et al. 2005b).}

We have compared the global properties of the groups analyzed here with the
Epic-pn camera and the global properties of these groups in the
literature. Complicating this comparison is that we have included the
central group galaxy in our global properties and we have not matched our
extraction radii to other studies. Despite this fact, we find { an agreement
within the statistical errors} between our global temperatures and the ROSAT
measurements given in Mulchaey et al. (2003) for groups with temperatures
below 1 keV. There are two groups above 1 keV in our sample and the XMM
results show that these groups have 20\% higher temperatures than reported
in the literature. A similar trend has been noted for groups studied with
ASCA (Hwang et al. 1999; Mulchaey et al. 2003). { One of these groups, NGC
5044, has been analyzed in detail by Finoguenov \& Ponman (1999), and a
comparison of the temperature profile has been made using ASCA and ROSAT
data. This analysis shows that the data agree outside the central
$30h_{70}^{-1}$ kpc. This possibly indicates an importance of
low-temperature components in defining the mean temperatures obtained with
ROSAT PSPC. Our temperature value of $1.21\pm0.01$ keV for NGC5044, obtained
within the $0.1-0.3r_{500}$ is in good consistency with the value of
$1.24\pm0.04$ keV, used in the M--T relation of Finoguenov et al. (2001).}

Detailed measurements of the temperature and element abundance distribution
in the studied systems allowed us to provide unbiased mean values for a
range of group properties which we list in Tab.\ref{t:prop}.  Column (1)
lists the system name, (2) the Galactic $N_H$, which is fixed in the spectral
modeling (3) the assumed luminosity distance, (4) implied scaling, (5)
the temperature, (6) metalicity as a fraction of the solar photospheric value,
derived from a single region fit with corresponding luminosity in the
0.5--2.4 keV band listed in (7). Tab. \ref{t:prop1}--\ref{t:prop2} list
name of the system (1), mass average temperature (2), metalicity (3),
volume averaged entropy (4) and pressure (5) using intervals $<0.1 r_{500}$
and $0.1-0.3r_{500}$, respectively.

\begin{table}
\begin{center}
\renewcommand{\arraystretch}{0.9}\renewcommand{\tabcolsep}{0.05cm}
\caption{\footnotesize
\centerline{Properties of groups within central 30$h_{70}^{-1}$ kpc.}
\label{t:prop}}

\begin{tabular}{rccccccccccc}
\hline
\hline
 Name & $N_H$     & $D_L$ & $\prime$ & kT & Z & $L_x$\\
      & $10^{20}$ & Mpc   &kpc& keV & $Z_\odot$&$10^{41}$\\
      &  cm$^{-2}$ &       &   &  &  &ergs/s\\
\hline
NGC~2300&5.27&28.0&8.05&$0.79\pm0.01$&$0.40\pm0.04$&1.03\\
NGC~3665&2.06&29.8&8.54&$0.40\pm0.02$&$0.25\pm0.08$&0.29\\
 IC~1459&1.18&24.2&6.94&$0.59\pm0.01$&$0.46\pm0.14$&0.53\\
NGC~3923&6.21&25.0&7.19&$0.57\pm0.01$&$0.20\pm0.01$&1.32\\
NGC~4168&2.56&32.8&9.40&$0.76\pm0.05$&$0.08\pm0.03$&0.19\\
NGC~4261&1.55&32.4&9.28&$1.24\pm0.03$&$0.33\pm0.04$&0.63\\
NGC~4636&1.81&13.3&3.86&$0.69\pm0.01$&$0.51\pm0.01$&8.14\\
NGC~5044&4.93&38.9&11.1&$0.90\pm0.01$&$0.56\pm0.01$&28.0\\
NGC~5322&1.81&25.4&7.31&$0.30\pm0.01$&$0.09\pm0.03$&0.08\\
NGC~5846&4.26&24.6&7.07&$0.70\pm0.01$&$0.45\pm0.01$&3.05\\
 \hline
\end{tabular}
\end{center}

\begin{center}
\renewcommand{\arraystretch}{0.9}\renewcommand{\tabcolsep}{0.05cm}
\caption{\footnotesize
\centerline{Properties of groups within $0.1 r_{500}$.}
\label{t:prop1}}

\begin{tabular}{rccccccccccc}
\hline
\hline
 Name & kT  & Z & S & P \\
      & keV & $Z_\odot$&keV cm$^2$ & $10^{-12}$ dyne cm$^{-2}$ \\

\hline
NGC~2300&$0.90\pm0.01$&$0.22\pm0.02$&$ 84\pm 4$&$ 1.77\pm0.10$& \\
NGC~3665&$0.54\pm0.16$&$0.26\pm0.31$&$ 44\pm21$&$ 1.45\pm0.70$& \\
 IC~1459&$0.61\pm0.02$&$0.06\pm0.01$&$ 25\pm 2$&$ 3.87\pm0.34$& \\
NGC~3923&$0.58\pm0.13$&$0.27\pm0.26$&$ 85\pm34$&$ 0.65\pm0.25$& \\
NGC~4168&$0.72\pm0.06$&$0.08\pm0.04$&$ 92\pm19$&$ 0.78\pm0.18$& \\
NGC~4261&$1.32\pm0.02$&$0.21\pm0.02$&$115\pm 4$&$ 2.75\pm0.11$& \\
NGC~4636&$0.79\pm0.00$&$0.26\pm0.01$&$ 58\pm 1$&$ 2.19\pm0.04$& \\
NGC~5044&$1.01\pm0.01$&$0.73\pm0.03$&$ 36\pm 1$&$ 10.3\pm0.3$& \\
NGC~5322&$0.64\pm0.09$&$0.06\pm0.06$&$ 56\pm22$&$ 1.71\pm0.65$& \\
NGC~5846&$0.86\pm0.01$&$0.38\pm0.02$&$ 61\pm 2$&$ 2.77\pm0.09$& \\
 \hline
\end{tabular}
\end{center}

\begin{center}
\renewcommand{\arraystretch}{0.9}\renewcommand{\tabcolsep}{0.05cm}
\caption{\footnotesize
\centerline{Properties of groups between $0.1 r_{500}$ and $0.3 r_{500}$.}
\label{t:prop2}}

\begin{tabular}{rccccccccccc}
\hline
\hline
 Name & kT  & Z & S & P \\
      & keV & $Z_\odot$&keV cm$^2$ & $10^{-12}$ dyne cm$^{-2}$ \\

\hline
NGC~2300&$0.75\pm0.01$&$0.48\pm0.19$&$161\pm19$&$ 0.39\pm0.07$& \\
 IC~1459&$0.59\pm0.03$&$0.09\pm0.03$&$108\pm19$&$ 0.41\pm0.08$& \\
NGC~4168&$0.77\pm0.31$&$0.00\pm0.05$&$155\pm73$&$ 0.44\pm0.24$& \\
NGC~4261&$1.11\pm0.02$&$0.17\pm0.01$&$164\pm 6$&$ 0.98\pm0.04$& \\
NGC~4636&$0.77\pm0.01$&$0.15\pm0.01$&$ 86\pm 3$&$ 1.10\pm0.05$& \\
NGC~5044&$1.21\pm0.01$&$0.38\pm0.02$&$ 86\pm 2$&$ 3.31\pm0.09$& \\
NGC~5846&$0.69\pm0.01$&$0.19\pm0.03$&$116\pm13$&$ 0.54\pm0.05$& \\
 \hline
\end{tabular}
\end{center}
\end{table}

We use the information derived from the maps as well as using annuli to
examine the radial variation of the properties of the hot gas. In
Fig.~\ref{cf:ent} we show the entropy and pressure profiles. These profiles
have been scaled using $r_{500}$ as calculated in Tab.\ref{t:b}, so that the
profiles are on a common scale. The predicted entropy for the groups is
shown as a dashed line and is derived from the scaling relationship of
Ponman et al. (2003). To scale the data we use $(T_w/10 keV)^{-2/3}$, where
T$_w$ is mass-weighted temperature measured within $0.1-0.3 r_{500}$. We
also examine the differences between the pressure profiles of the groups in
our sample. As entropy, $T n^{-2/3}$, scales as $T_w^{2/3}$, the density
scales as $T_w^{1/2}$ and thus the pressure scales as $T n \sim
T_w^{3/2}$. We then apply this scaling to the data, using $(T_w/10
keV)^{-3/2}$, and display a typical pressure profile of the cluster scaled
the same way in Fig.~\ref{cf:ent}.

The increase in entropy with increasing radius is roughly proportional to
$r^{1.1}$.  The adopted scaling is insensitive to the choice of the $T_w$,
since it only adjusts the points parallel to the scaling. This is the
advantage of entropy scaling. { In Finoguenov et al. (2005a) it is shown
that clusters with cool cores follow the $r^{1.1}$ law and including all
clusters flattens the trend. In addition to the clusters we have included
the entropy and pressure profiles, obtained using the high-redshift
subsample of Mulchaey groups (Finoguenov et al. 2005b) to extend the result
down to group scales. In figure 3 the only group entropy profile that
follows the $r^{1.1}$ between $0.1<r_{500}<0.3$ is the NGC~5044 group.  In
the right entropy panel the NGC~5846 group lies above the scaling relation
and the NGC~4636 group data does not extend to sufficiently large radius to
determine if it follows the scaling relation. In the right entropy panel
only the NGC~2300 group might follow the entropy scaling at $r_{500}>0.1$
but the outer most point is not consistent with the $r^{1.1}$ scaling.
Looking at the solid lines in figure 3 for the group entropy profiles from
Finoguenov et al. (2005a), our low redshift sample follows that scaling much
more closely. However, the data for NGC~5044 and NGC~4636 lie systematically
below the line for the average entropy profile while in the right panel the
groups have a higher average entropy than the Finoguenov et al. (2005a)
sample. One can further subdivide the groups based on those that have higher
entropy at the center (e.g. NGC~4261) and those that have higher entropy at
the outskirts (e.g. NGC~5322), compared to the plotted scaling relations. In
both cases systematic deviations are also present in the pressure profiles.}

The NGC4636 group shows a higher pressure, compared to the mean, as also
found in other cool core systems (Finoguenov et al. 2005a).  NGC~5044,
NGC~5846 and NGC~4168, show pressure profiles consistent with the mean. The
profiles of NGC~4261, NGC~2300 and IC~1459 initially decline more rapidly,
but then become flatter and start to match the scaling for pressure. The
remaining three groups show pressure profiles that decline more quickly than
the typical profile. The significance of this can be seen by considering the
response of the pressure to the large-scale state of the gas and
gravitational potential, where $M/r\propto T d lg(P) / d lg(r)$ and $P
\propto M$, since both P and M scale with $T^{3/2}$. Lower pressure either
indicates that the gravitational mass is low or that the temperature is
enhanced at even larger radii. In the former case, the total mass at
$r_{500}$ should be $1-5\times10^{12} M_\odot$. In the latter case, some
additional entropy input is required in excess of that predicted by the
modified entropy scaling. This additional entropy is needed on scales
exceeding $0.3 r_{500}$ (approximately 150 kpc in these groups, see
Fig.~\ref{cf:ent}).

Another way to look at the results described above is to present the
normalization and the slope of the entropy and pressure profiles
(Tab.\ref{t:fits}). To deal with the non-statistical scatter of the points,
we applied the bisector method and orthogonal regression, following Helsdon
\& Ponman (2000) and present the results obtained using the latter method in
Tab.\ref{t:fits}, while monitoring the difference between the methods. Once
the general behavior of the entropy and pressure is found, we studied the
amplitude of the deviations from the fit and compared that with the
statistical errors in Tab.\ref{t:scatter} for the systems with the best
statistics.  In addition to fitting the power laws to the pressure and
entropy profiles, we provide in Tab.\ref{t:scatter} the scatter around the
mean trend for the groups from Finoguenov et al. (2005b), where we allowed a
renormalization of the model. In general, the structure in both entropy and
pressure maps amounts to $10-20$\%. Stronger deviations, seen in NGC~2300,
are located in the area between the two major galaxies, indicating an
interaction between the two dominate galaxies in this group.  The relatively
high level of fluctuations in NGC~4261, discussed in more detail below, seem
to be associated with the AGN activity.

\begin{table}
\begin{center}
\renewcommand{\arraystretch}{0.9}\renewcommand{\tabcolsep}{0.05cm}
\caption{\footnotesize
\centerline{Characteristics of entropy and pressure profiles for the sample.}
\label{t:fits}}

\begin{tabular}{rccccc}
\hline
\hline
 \multicolumn{1}{c}{Name} & S($0.2r_{500}$) &   S slope  &  P($0.2r_{500}$) &   P slope\\
      & keV cm$^2$ & & \multicolumn{2}{l}{$10^{-14}$ dyne cm$^{-2}$} \\
\hline
NGC~2300&$163\pm10$& $0.80\pm0.10$& $46\pm2$& $-1.17\pm0.08$ \\
NGC~3665&$116\pm80$& $0.64\pm0.48$& $ 8\pm4$& $-1.58\pm 0.44$\\ 
 IC~1459&$136\pm 6$& $0.81\pm0.05$& $46\pm5$& $-1.08\pm0.20$ \\
NGC~3923&$210\pm20$& $1.07\pm0.17$& $17\pm5$& $-1.44\pm0.26$ \\
NGC~4168&$ 90\pm10$& $0.20\pm0.13$& $50\pm8$& $-0.72\pm0.18$  \\
NGC~4261&$288\pm32$& $0.87\pm0.15$& $95\pm6$& $-0.98\pm0.14$ \\
NGC~4636&$131\pm 5$& $0.81\pm0.03$& $66\pm1$& $-1.07\pm0.04$ \\
NGC~5044&$104\pm 2$& $0.82\pm0.04$& $284\pm13$& $-0.78\pm0.07$   \\
NGC~5322&$337\pm23$& $0.77\pm0.26$& $6.5\pm2$& $-1.29\pm0.34$  \\
NGC~5846&$116\pm 6$& $0.87\pm0.05$& $58\pm2$& $-1.26\pm0.05$ \\
\hline
\end{tabular}
\end{center}

\begin{center}
\renewcommand{\arraystretch}{0.9}\renewcommand{\tabcolsep}{0.05cm}
\caption{\footnotesize
\centerline{Entropy and pressure fluctuations around the best-fit, per cent.}
\label{t:scatter}}

\begin{tabular}{rccccccccccc}
\hline
\hline
 Name & $\sigma$S, \%  &   $\sigma$P, \%  & $\sigma$S, \%  &   $\sigma$P, \%
 & Comment\\
& \multicolumn{2}{c}{power law} &\multicolumn{2}{c}{averaged group profile}\\
\hline
NGC~2300&$25\pm 6$&$126\pm15$ & $24\pm 6$ & $171\pm 22$\\
NGC~3923&$31\pm 8$& $25\pm12$ & $18\pm12$ & $36\pm 8$\\
NGC~4261&$21\pm 2$& $17\pm 2$ & $15\pm 1$ & $20\pm 2$& \\
NGC~4636&$24\pm 1$& $22\pm 2$ & $20\pm 1$ & $21\pm 2$&including the plume\\
NGC~4636&$11\pm 1$& $23\pm 3$ & $16\pm 1$ & $31\pm 2$ &central  $5^\prime$\\
NGC~5044&$10\pm 2$& $13\pm 2$ & $12\pm 3$ & $8\pm 2$& central $5^\prime$\\
NGC~5846&$13\pm 2$& $16\pm 3$ & $14\pm 1$ & $12\pm 3$&\\
\hline
\end{tabular}
\end{center}

\begin{center}
\renewcommand{\arraystretch}{0.9}\renewcommand{\tabcolsep}{0.05cm}
\caption{\footnotesize
\centerline{Fe abundance trends and scatter.}
\label{t:fe}}

\begin{tabular}{rccccccccccc}
\hline
\hline
 Name & Fe ($0.05r_{500}$) & Fe slope & $\sigma$Fe, \% & Comment\\
\hline
NGC~4636&$0.46\pm 0.06$ & $-0.4\pm0.2$& $45\pm 6$  &central  $5^\prime$\\
NGC~5044&$0.68\pm 0.06$ & $-0.3\pm0.1$& $25\pm 6$ & central  $5^\prime$\\
NGC~5846&$0.34\pm 0.09$ & $-0.9\pm0.3$& $41\pm 6$  &\\
\hline
\end{tabular}
\end{center}
\end{table}

\subsection{Origin of the X-ray emission}

The XMM-Newton data allow us to compare the entropy and pressure behavior at
radii exceeding $0.1 r_{500}$ for a large sample of groups for the first
time.  Based on the levels of entropy and pressure plotted in
Fig.\ref{cf:ent} which are further confirmed in normalization of the
pressure in Tab.\ref{t:fits}, we conjecture a subdivision of our sample onto
two broad categories. We base our division on X-ray properties at
$0.2r_{500}$. This results in groups, whose properties are largely defined
by gas accretion (ADAGE - accretion dominated average group environment) and
systems whose diffuse thermal X-ray emission is not wholly associated with
the accretion of the gas on potential wells of the group. The first category
is also speculated to be baryonically closed groups based on the
extrapolations of Mathews et al. (2005). Inside $0.2r_{500}$, we report
large deviations in the mean values for entropy and pressure between the
ADAGE groups (Tab.\ref{t:fits}), which suggests further refinement of the
definition. We call NGC~5044, NGC~5846 regular groups (or cooling core
groups), because of low scatter in their pressure and entropy profiles. In
addition, these systems follow the expected entropy scaling relation, shown
in Fig.\ref{cf:ent}. Although, the central properties of NGC~4636 also
follow the scaling relations, large deviations are seen at $0.2r_{500}$
radius, which we attribute to an interaction with the Virgo cluster.  Many
underluminous ADAGE groups, are similar to regular groups at large
radii. These are IC~1459, NGC~2300, NGC~4168 and NGC~4261. For NGC~2300 and
NGC~4261, for which the two dimensional analysis was possible, we measure a
large degree of disturbance in their gas properties. NGC~2300 might be
experiencing a merger, while NGC~4261 has signatures of ongoing AGN
activity.

To the second category of objects belongs NGC~5322, and perhaps NGC~3923
based on our failure to detect emission at $0.2r_{500}$. Although,
there is a large uncertainty in parameters derived for the  NGC~3665 group, it
exhibits extremely low galaxy velocity dispersion for its temperature and we
include it into this category.

As the division is entirely based on the entropy level, a system of low
X-ray luminosity is considered a group if its temperature is proportionally
low, while a different origin of the emission is suggested if the
temperature is high. With the current data in hand, these two categories do
not overlap in the entropy at radii exceeding $0.3 r_{500}$.

ADAGE groups are virialized objects, whose thermodynamic properties of the
X-ray emitting gas have been determined by the accretion of gas into the
potential well of the group.  Arguments on the conversion of neutral HI gas
to hot X-ray emitting gas (Trinchieri et al. 2003) indicate that the nature
of X-ray emission in the compact group HCG~92 is different.  The importance
of defining ADAGE is that a similar approach is adopted for simulations
(e.g. Borgani et al. 2005). The weakest point, shared by all X-ray studies
of groups, is that the low-mass clusters and groups show substantial
deviations from the scaling laws predicted by pure gravity. So, a common
approach in the modeling is to tune various feedback effects so that the
simulations reproduce the observed X-ray properties of the groups. A
subdivision, attempted in this paper, is aimed at outlining the potentially
different origin of the X-ray emission of a subclass of X-ray underluminous
groups , which is usually considered in simulations looking into the
formation of early-type galaxy (e.g. Burkert \& Naab 2004), while on the
other hand producing a sample of objects whose properties could be compared
to cluster-type simulations of e.g. Borgani et al. (2004).

\begin{figure*}

\includegraphics[width=6.in]{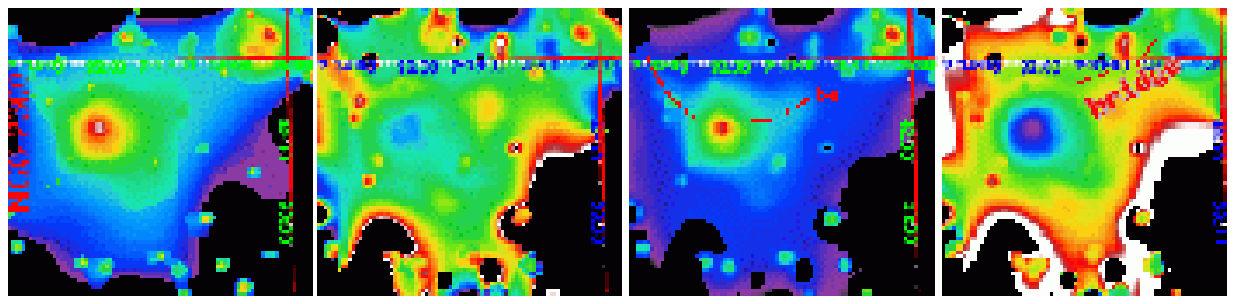}

\includegraphics[width=6.in]{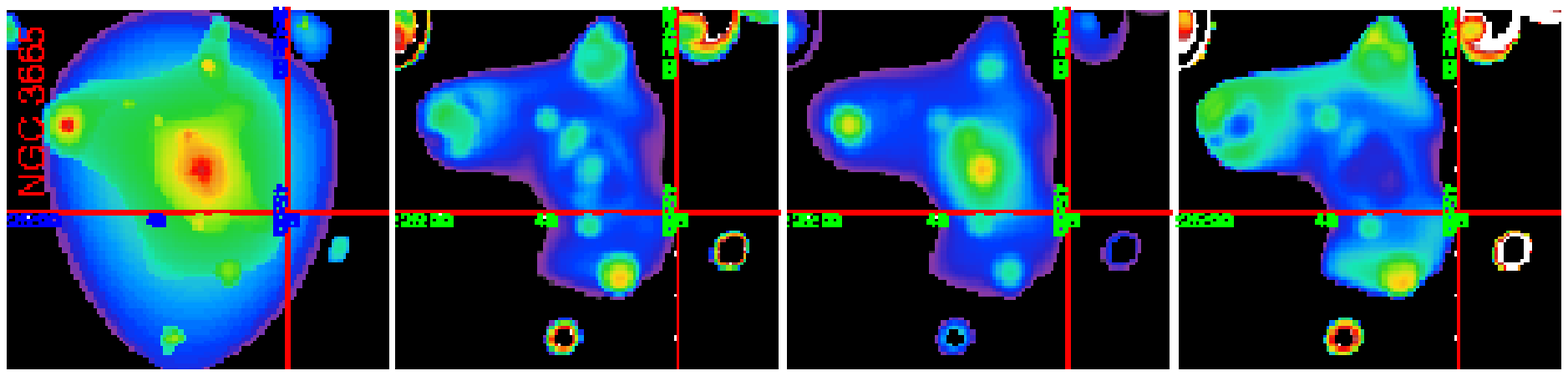}

\includegraphics[width=6.in]{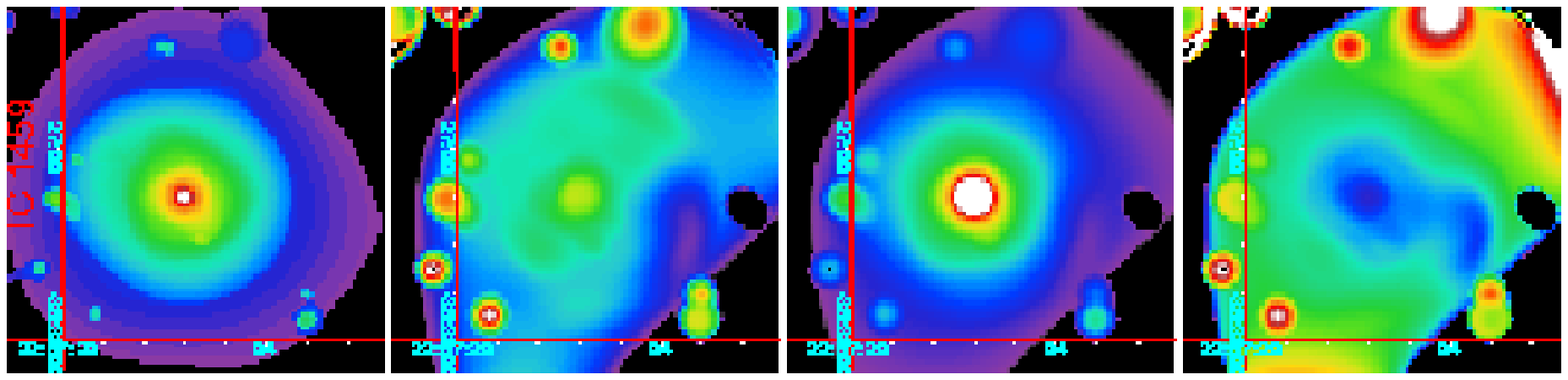}

\includegraphics[width=6.in]{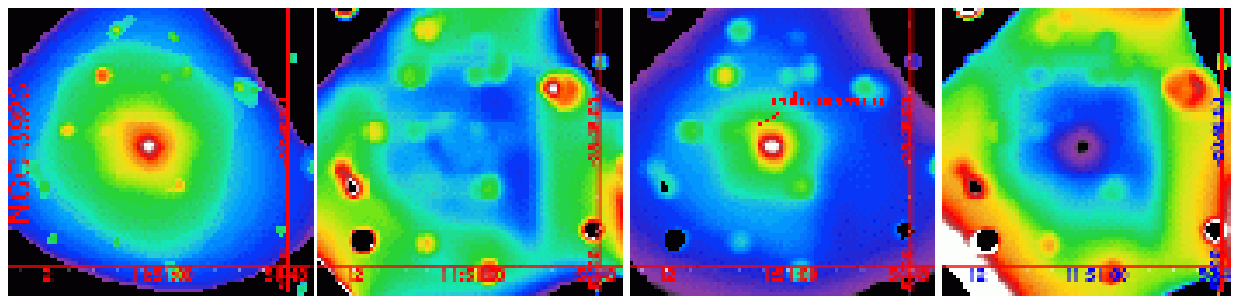}

\includegraphics[width=6.in]{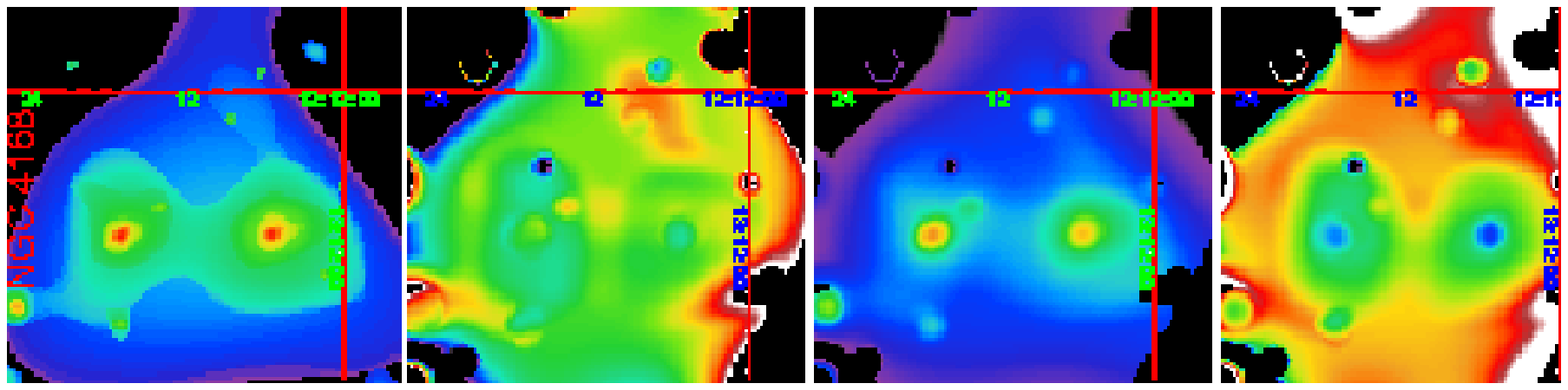}
\end{figure*}

\begin{figure*}
\includegraphics[width=6.in]{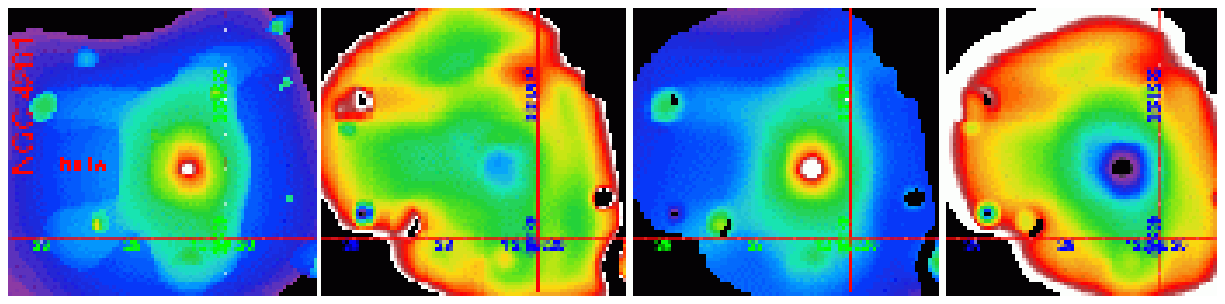}

\includegraphics[width=6.in]{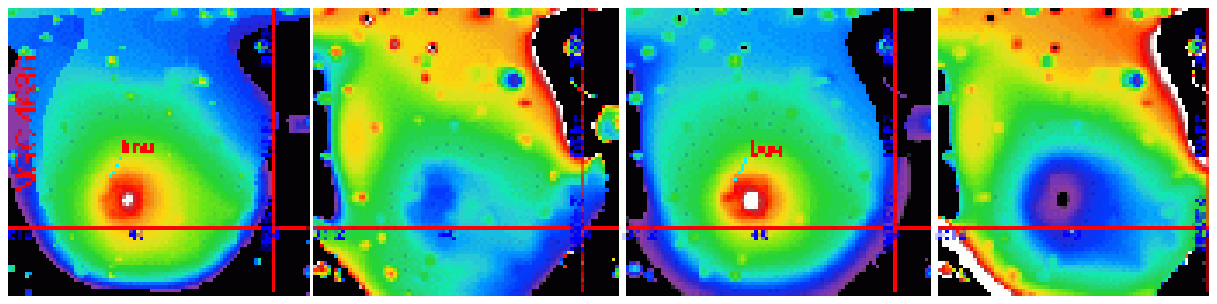}

\includegraphics[width=6.in]{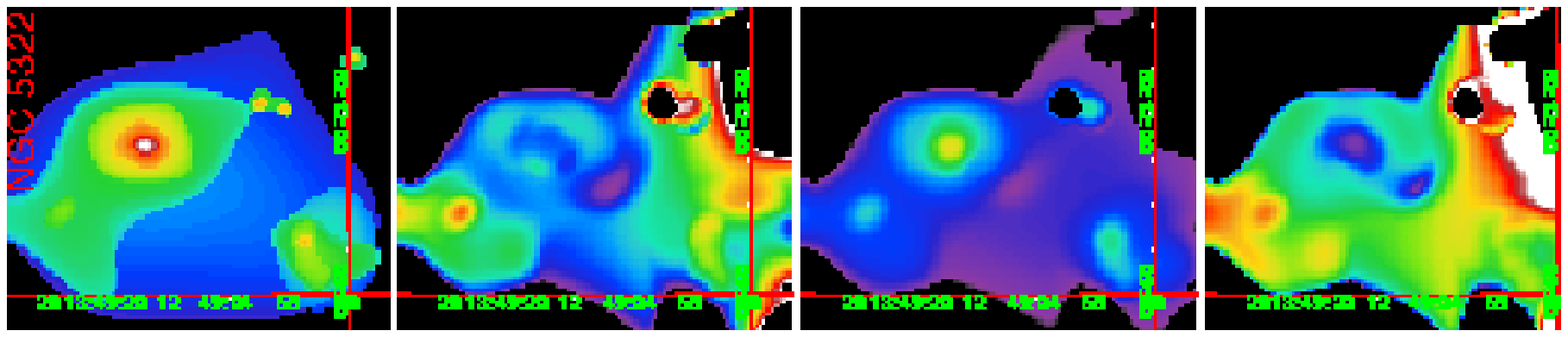}

\includegraphics[width=6.in]{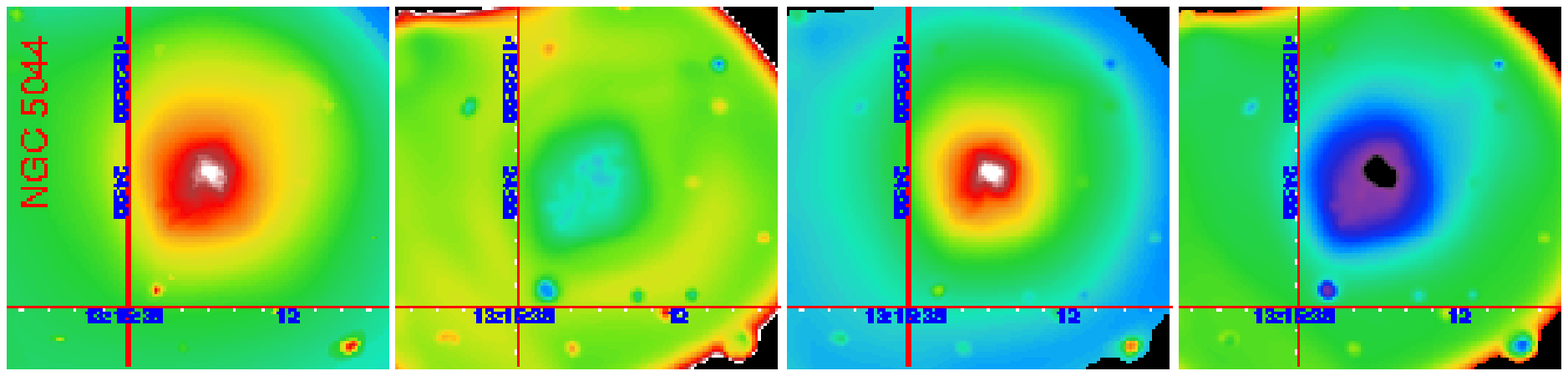}

\includegraphics[width=6.in]{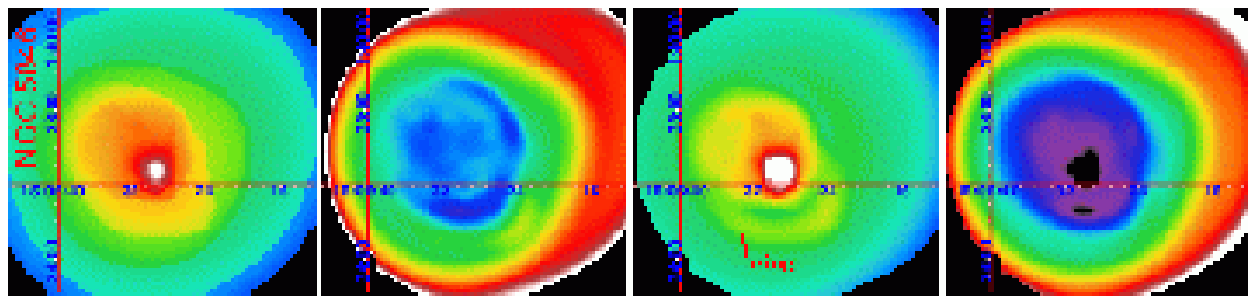}

\figcaption{From left to the right: surface brightness, temperature,
pressure and entropy maps of the groups. From top to bottom the panels
correspond to NGC~2300, NGC~3665, IC~1459, NGC~5322, NGC~4168; NGC~4261,
NGC~4636, NGC~5322, NGC~5044, NGC~5846.The temperature, pressure and entropy
maps in this figure are based on the hardness ratio analysis.  The values
for the pseudo entropy and pressure maps are arbitrary, while their absolute
values as well as the significance of all the features are obtained through
the subsequent spectral analysis, which we both tabulate and report in the
text. The features not discussed in the text shall be regarded as marginal.
All the figures have the same color scale for the same quantity. Temperature
coding is 0.3 -- violet, 0.5 -- blue, 0.7 -- cyan, 0.9 -- green, 1.2 --
yellow, 1.6 -- red. On the panel corresponding to the pressure image of
NGC~2300, red line demarcates the bow shaped enhancement, the bridge in the
entropy map of NGC2300 is labeled. Enhancement in the pressure map of
NGC3923 is marked. The bow on the images of NGC4636 is marked. On the panel
corresponding to the pressure image of NGC~5846, red line demarcates the
ring shaped enhancement. The levels of the surface brightness could be read
from Fig.\ref{f:ol}.
\label{cf:1}}

\end{figure*}

\begin{figure*}
\vspace*{-1.5cm}

\mbox{
\includegraphics[width=6.cm]{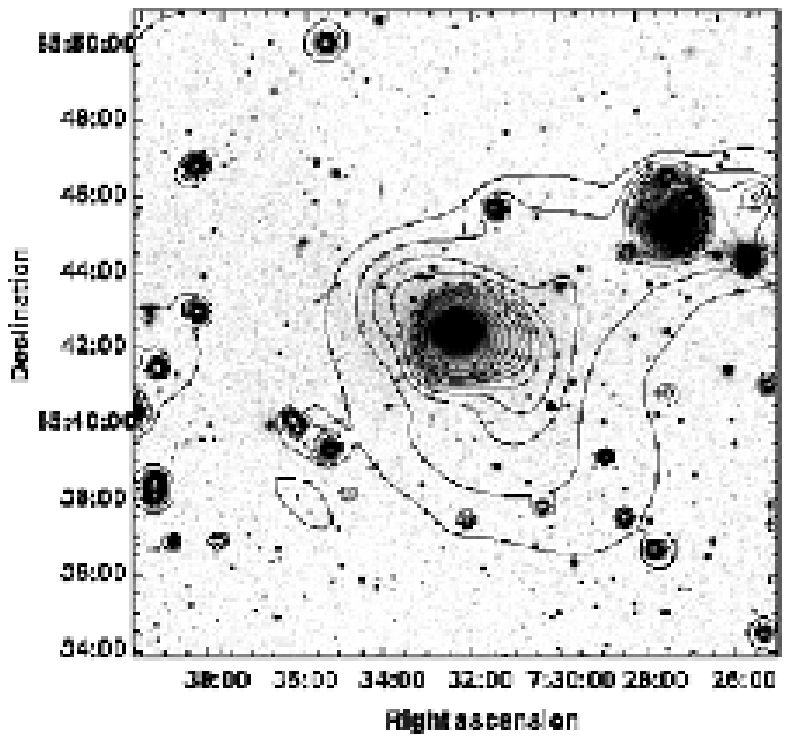}
\includegraphics[width=6.cm]{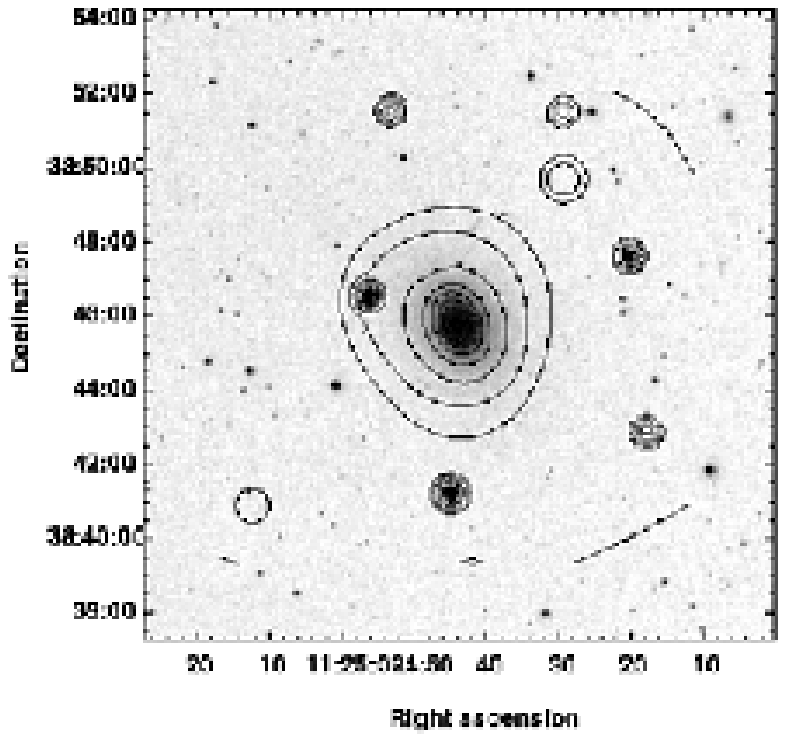}
\includegraphics[width=6.cm]{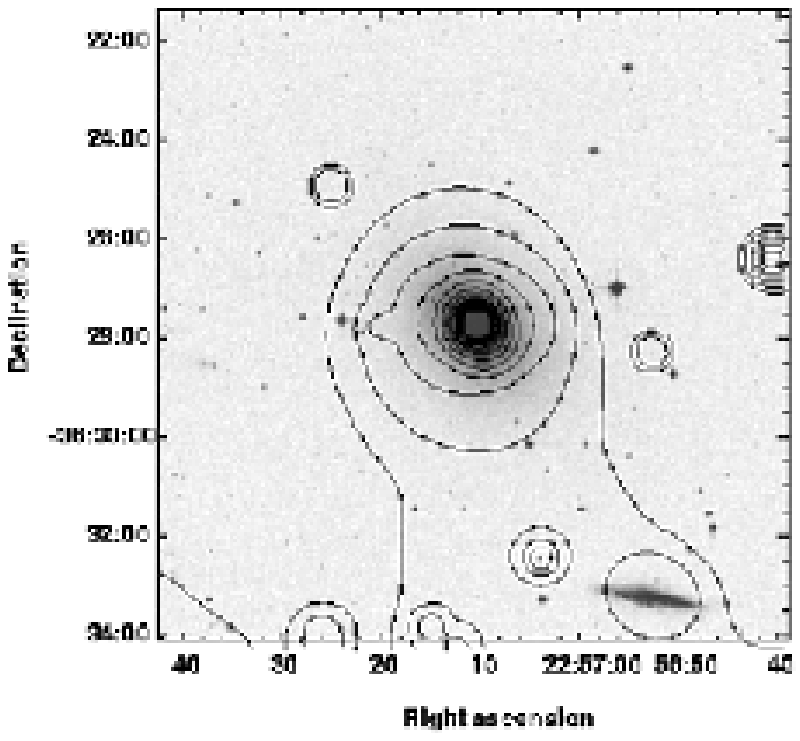}}

\mbox{ \includegraphics[width=6.cm]{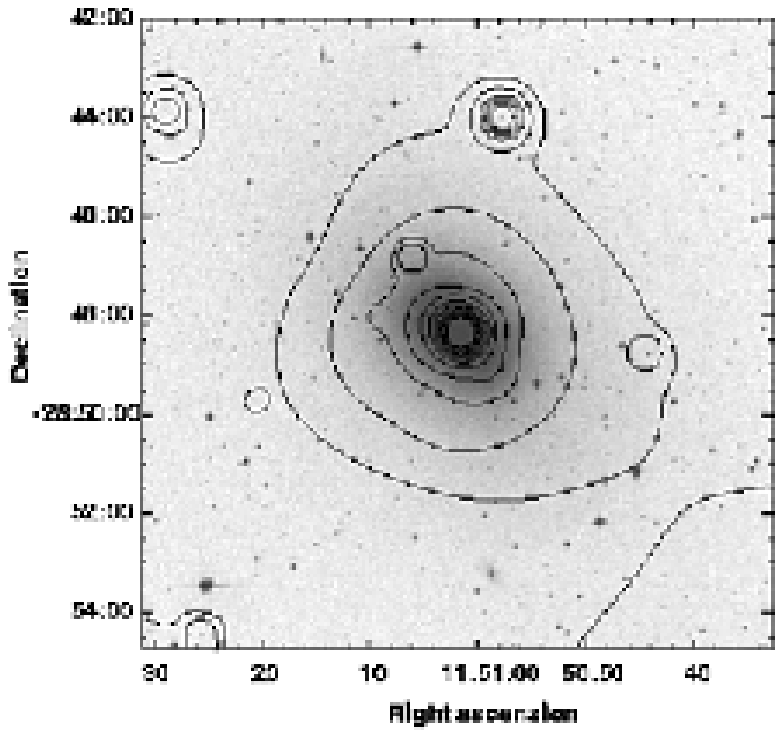}
\includegraphics[width=6.cm]{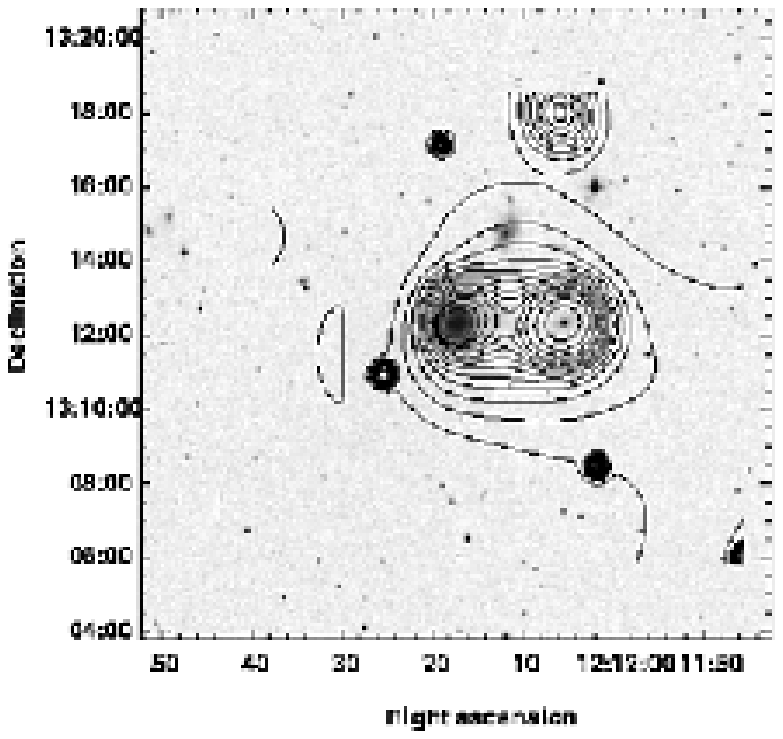}
\includegraphics[width=6.cm]{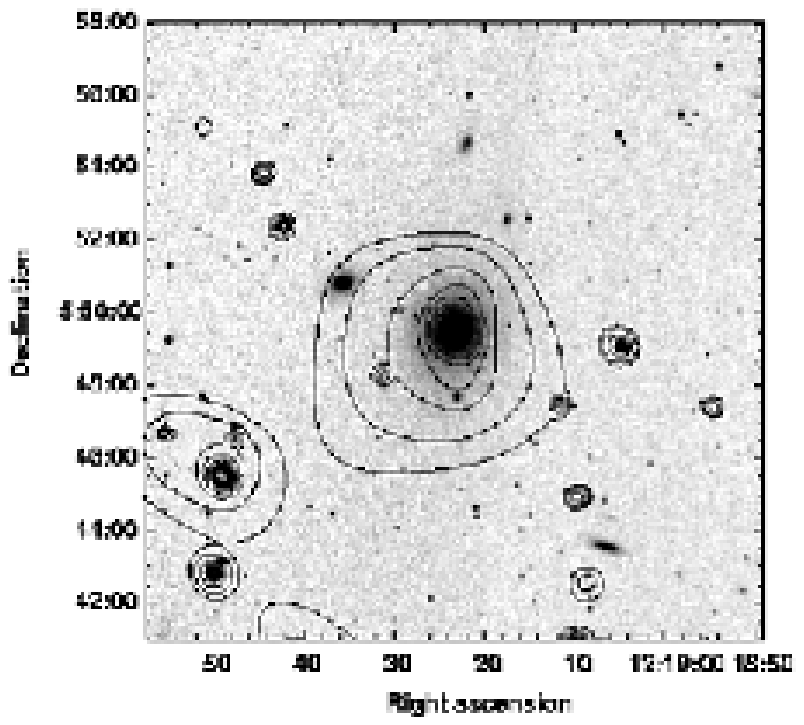}}

\mbox{ \includegraphics[width=6.cm]{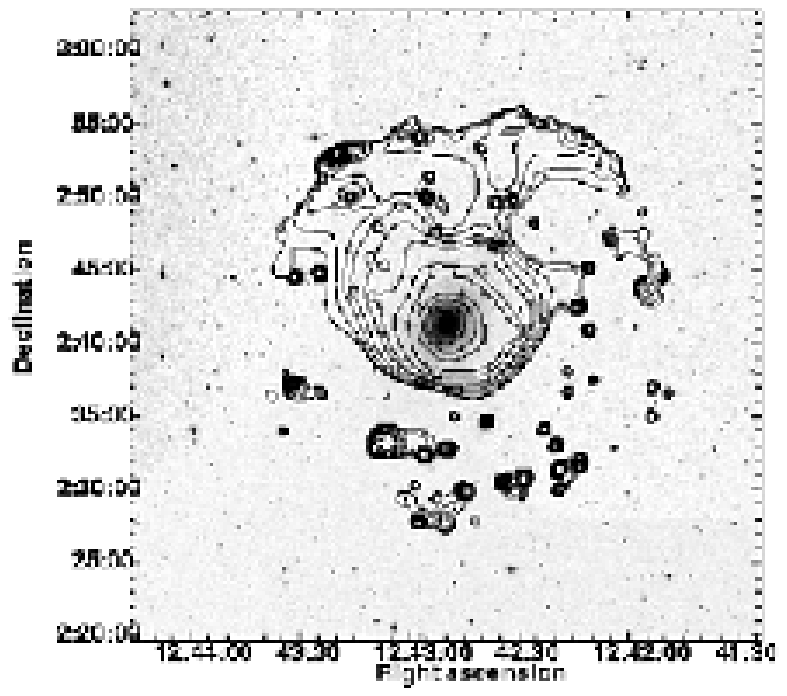}
\includegraphics[width=6.cm]{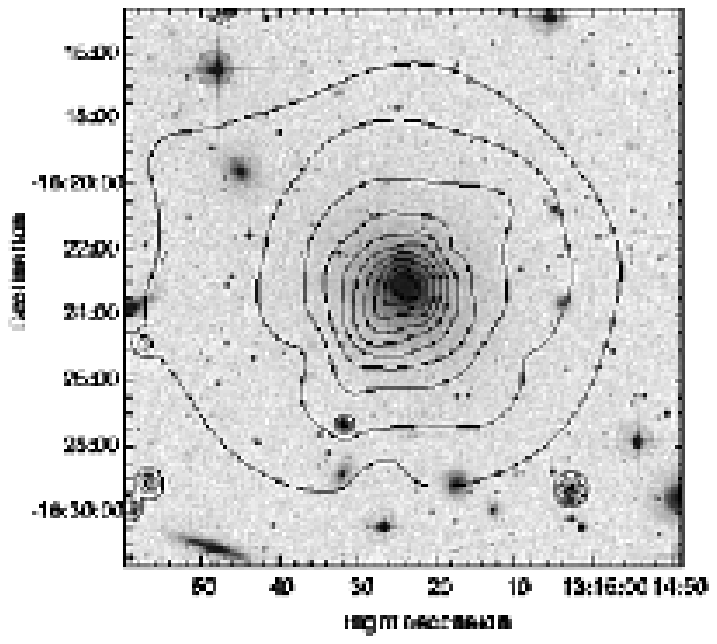}
\includegraphics[width=6.cm]{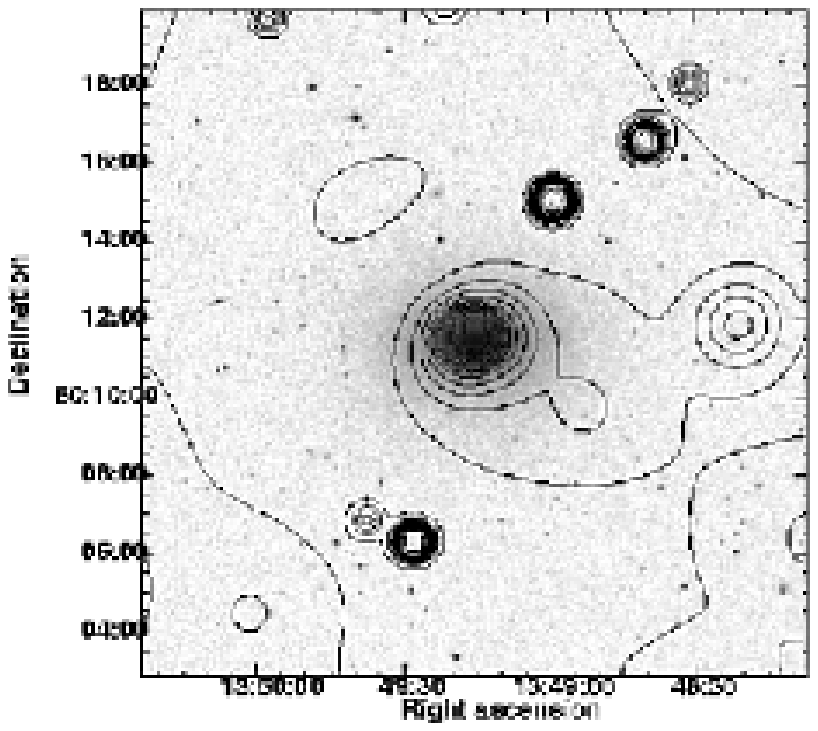}}

\mbox{ \includegraphics[width=6.cm]{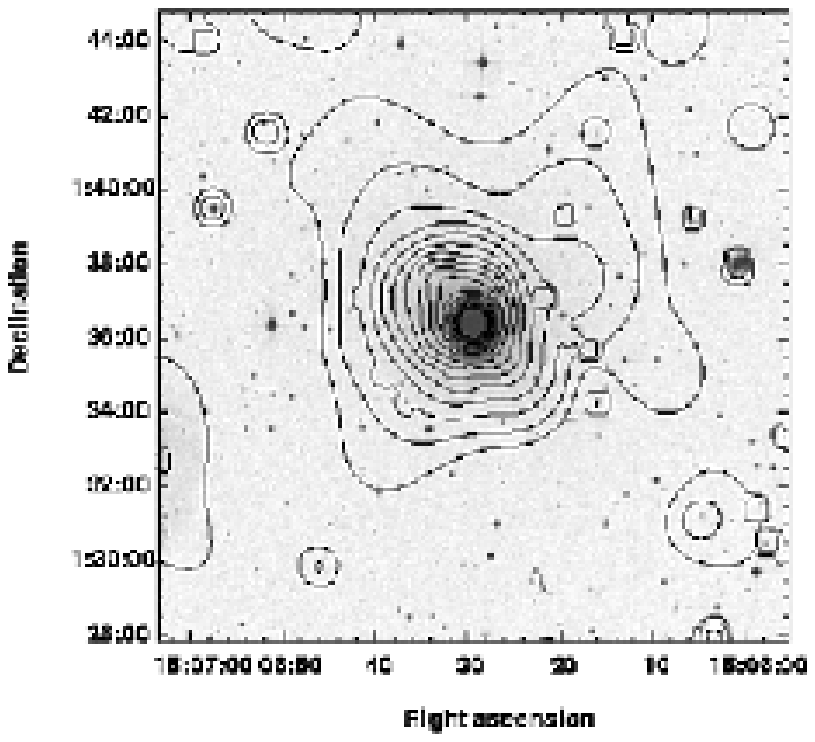}
\includegraphics[width=6.cm]{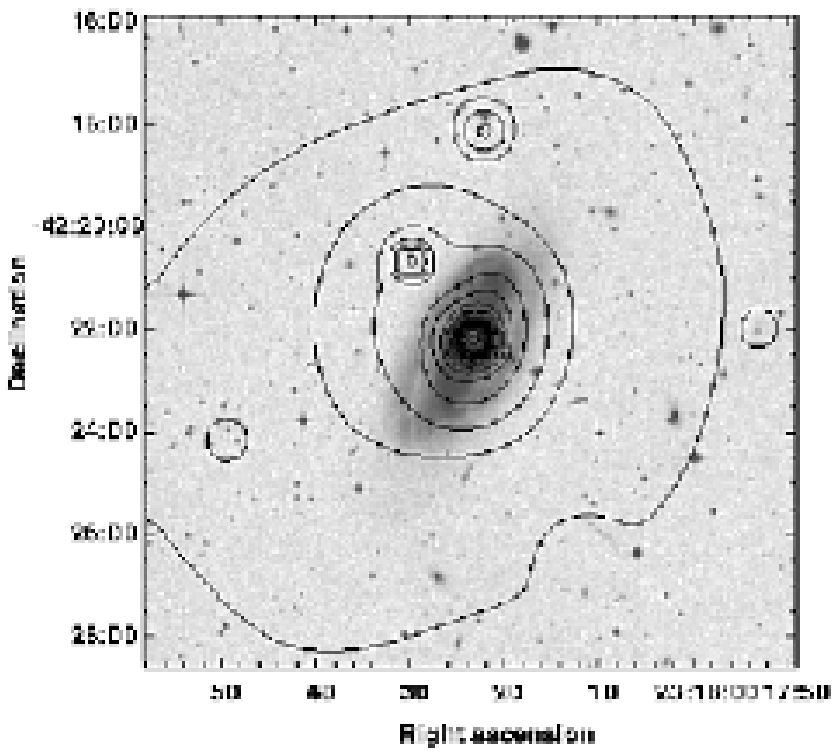}}

\figcaption{\footnotesize DSS2 R band images of the groups overlaid with
contours indicating levels of X-ray emission, as obtained by the adaptive
smoothing. From upper left to lower right NGC~2300, NGC~3665, IC~1459,
NGC~3923, NGC~4168, NGC~4261, NGC~4636, NGC~5044, NGC~5322, NGC~5846,
NGC~7582 are shown.  The X-ray contours are in units of MOS1 count/pixel
(pixels are $4^{\prime\prime}$ on a side) and drawn on the following levels:
0.2-0.4-0.5-0.6-0.7-0.8-0.9-1.1-1.3-1.4-1.7-2.1-3.2-4.8 (NGC~2300);
0.10-0.12-0.2-0.3-0.5-0.8-1.0-1.4-1.8-2.3-2.8-3.4-4.0-4.6-5.4 (NGC~3665);
0.1-0.2-0.4-0.7-1.1-1.6-2.3-3-4-5-6-7-9-10-12-13-15 (IC~1459);
0.2-0.3-0.8-2-3-4-6-8-10 (NGC~3923);
0.08-0.10-0.13-0.15-0.17-0.19-0.22-0.24-0.26-0.29-0.32-0.35-0.39-0.6
(NGC~4168); 0.22-0.27-0.3-0.4-0.6-1.1-1.4-2-3-4-5-7-9-11 (NGC~4261);
0.87-0.95-1.0-1.2-1.4-1.7-2-3-6-9-19-39-57-116 (NGC~4636);
0.3-0.6-1.4-3-5-7-10-14-18-23-28 (NGC~5044);
0.04-0.06-0.07-0.08-0.10-0.12-0.14-0.20-0.35-0.5-0.9 (NGC~5322);
0.2-0.3-0.7-1.3-2-3-5-6-8-10-13-15-18-21-24-28 (NGC~5846);
0.16-0.17-0.19-0.21-0.25-0.30-0.35-0.42-0.5-0.6-0.7-0.8 (NGC~7582).
In each figure the lowest contour level is selected to be $\sim$4$\sigma$
above the background. 
\label{f:ol}}
\end{figure*}

The entropy levels of the second category of objects are up to a factor of 3
higher than the other groups in our sample, as illustrated with NGC~5322 in
Fig.\ref{cf:ent}, and the ASCA sample of Ponman et al. (2003).  The low
pressure (up to a factor of 10) and high entropy at all radii in these
systems cannot be explained by infall of gas associated with group
formation. Thus, it is probably more likely that the X-ray emission in these
systems originates from the ISM of the galaxies and not as infalling gas.
We can infer by looking at the normalization of the pressure profiles that
most of the systems in question have masses about 10\% of the NGC~5044
group, which places them into $10^{12}$ M$_\odot$ range. This is more
typical of massive elliptical galaxies. The existence of such objects
suggests the mere presence of a faint X-ray emission in a system can be
misleading. In particular, the presence of X-ray emission in a group does
not always mean the system is a virialized object and the nature of these
groups is still unclear. To complete the global properties of this category
of groups, we note that the element abundance observed in the X-ray gas is
very low, and that (under assumption of the robustness of optical metalicity
indicators) contradicts the suggestion that all the gas originates from
stellar mass loss.  Additional insights into the origin of this component
has been provided by the higher redshift part of this survey, which includes
HCG92 (Finoguenov et al. 2005b), for which Trinchieri et al. (2003) argue
that shock heating of an HI cloud is now generating the X-ray emitting group
gas. They reach the conclusion that the faint diffuse X-ray component is
from a galaxy interacting with the group. We note that NGC~5322 has a
counter-rotating core, which is most certainly a result of a recent
merger. The high entropy level of the observed component requires cooling
times longer than the Hubble time to remove the high entropy gas (e.g. Voit
et al. 2003).  So we expect the faint X-ray halo to be an indicator for and
a fossil record of the merger history of early-type galaxies.  These
arguments can also be extended to the faint X-ray halos of post-merger
galaxies, reported in Read \& Ponman (1998).  On the other hand, spiral
dominated groups studied here (NGC~7582) and in the literature (Mulchaey et
al. 1996) do not have detected extended X-ray emission on 100 kpc scales,
with first detections of extended emission on 20 kpc scales by Pedersen et
al. (2005).  This may indicate that the majority of spiral rich groups
neither possess large mass halos nor have undergone the galaxy-galaxy
interactions needed to form a diffuse hot halo, similar to X-ray faint
groups like NGC~5322 and HCG~92.

\subsection{Maps}

We present in this section the surface brightness, temperature,
projected pressure and entropy maps in Fig.\ref{cf:1}, where the
details of the map construction are described above. Overlays with
DSS2 R band images are shown in Fig.\ref{f:ol} and individual systems
are discussed in the subsections below. Although the maps present
projected properties, they could be shown to give an insight on the 3d
properties (Schuecker et al. 2004; Dolag et al. 2005).

\section{Discussion}

Before discussing the individual systems, we outline the sources of
variations seen in the maps in Fig.\ref{cf:1}. Among the four maps
presented, only two are independent and in principle sufficient for the
analysis. Although surface brightness analysis and temperature maps are more
commonly used in studies of groups and clusters, pressure and entropy maps
are more straight-forward to interpret. The pressure traces the generalized
stress tensor, which in the case of a system in hydrostatic equilibrium and
without dynamically important magnetic fields is given by the gravitational
potential.  Departures from equilibrium are likely to be local, i.e. seen as
pressure fluctuations on the map. Common examples include shocks
(e.g. Markevitch et al. 2002; Jones et al. 2002; Henry et al. 2004;
Finoguenov et al. 2004a, 2004b) and pressure waves (e.g. Fabian et al. 2003;
Schuecker et al. 2004). Substructure can also be seen in pressure maps as
secondary peaks, provided that it retains its own gas or it moves
subsonically. The entropy map for a system in equilibrium should be
symmetrical around the center and exhibit an increasing entropy level with
increasing radius. Deviations in the entropy maps in the form of low entropy
gas displaced from the center are either due to development of instabilities
(e.g. cold fronts in A3667, Vikhlinin et al. 2001 and in A3562, Finoguenov
et al. 2004b; shear instability in A3667, Mazzotta et al. 2002; Briel et
al. 2004), which are slow to relax (0.2 of the sound speed, Ricker \&
Sarazin 2001) or the result of stripping (M86, Finoguenov et
al. 2004a). These are examples of low entropy gas displaced from the
center. Off-center low entropy zones associated with local gravitational
minima (e.g. secondary pressure peaks, or simply galaxies, as in Finoguenov
et al. 2004c), are indicators of substructure.  In addition, areas of
high-entropy gas, found surrounding low-entropy gas (so called buoyant
bubbles) are produced by in situ heating, most likely by AGN (for examples
see NGC~4261 below).

Since the X-ray emission of groups is strongly dominated by Fe lines, the
maps are also affected by patchy metalicity distributions. In our discussion
below of individual features, we deconvolve the effects of enhanced density
from the effects of the enhanced metalicity or indicate where this
separation is not obvious. Patchiness of the metalicity distribution has not
been studied in detail before and could be an important factor indicating
the processes of the metal release into the intragroup medium. One of the
well-established examples of such a process is the gas entrainment by the
jet-induced instabilities in M87 (Churazov et al. 2001), which brings
low-entropy, high metalicity gas from the inner kpcs of M87 out to 50 kpc
radius.

The systematic analysis of the metalicity distribution { using the regions}
in NGC~4636, NGC~5044 and NGC~5846 is reported in Tab.\ref{t:fe}. { With the
typical size of the resolution element of 2--10 kpc, sampling the area
within 20--50 kpc from the brightest group galaxy}, we find that the
profiles are consistent with a linear decrease with radius, reaching 0.3
solar value by $0.05r_{500}$. The scatter of points around the best fit is
high, 30--50\%, which characterizes the degree of patchiness { on 20--50 kpc
scales}.

\subsection{NGC~2300, $z=0.0046$}

The pressure and entropy structure consists of two blobs, centered on
NGC~2300 and NGC~2276. The pressure map contain a large-scale bow-shape
enhancement outside the low entropy gas centered on NGC~2300. Such gas
compressions are possible in several cases: supersonic motion of NGC~2300
with respect to the group; supersonic interaction between NGC~2300 and
NGC~2276; presence of substructure around NGC~2300. In any case, the
dispersion in the pressure map of NGC~2300, reported in Tab.\ref{t:scatter},
is high, in comparison to other groups. Since this does not seem to be an
effect of confusion with a background object, as is sometimes the case
(Mahdavi et al. 2005), we suggest, based on a similar work applied to
simulations (Finoguenov et al. in prep.), that the NGC~2300 group exhibits
the signs of a merger. The pressure and entropy deviations $2^\prime$
southwest of the center may be explained if there is ongoing gas accretion
from the south. NGC~2276 (Arp 25) is considered a strongly interacting
object (Gruendl et al. 1993; Hummel \& Beck 1995), which supports our
suggestion on the supersonic interaction. The line-of-sight velocity
difference between NGC~2276 and NGC~2300 amounts to $503\pm7$ km s$^{-1}$,
which is supersonic with respect to the sound speed for the IGM temperature
of 0.7 keV. In general, late-type spiral galaxies are considered to be
currently infalling and moving on radial orbits (Biviano \& Katgert 2004).


There is an entropy bridge (shown in green color in Fig.\ref{cf:1}) toward the
second galaxy (NGC~2276), suggesting absence of a large line-of-sight
separation between the two.  This implies that gas in the halo of the spiral
galaxy is interacting with IGM of the group.

Spectroscopic results suggest that apart from the very center of the
elliptical and one arcminute around the spiral, the temperature distribution
is nearly isothermal. Extensions of the pressure to the south-west and to
the west are confirmed and are evaluated to be a factor $1.33\pm0.05$
enhancement. The extension of the low entropy zone to the south-west is
characterized by a $10\pm1$ \% lower entropy and the entropy bridge, which
has an entropy of $50\pm10$ keV cm$^2$, is a factor of 2 lower, both
compared to a typical entropy at the same distance to NGC~2300, estimated
from the annuli.

\subsection{NGC~3665, $z=0.0069$}

The bright part of the emission extends to $3^\prime$ in radius,
centered on NGC~3665 and has a 
similar elongation as the galaxy's major axis. There is also a very
bright background AGN east of the galaxy.

Due to lack of counts, the spectral analysis was done on 4 zones that
are symmetric around the center, which precludes revealing any 2d structure.

\subsection{IC~1459, $z=0.0057$}

IC~1459 is a well studied elliptical galaxy with a known hard X-ray component 
suggestive of an AGN (Matsumoto et al. 1997). Recent Chandra observations of
this system reveal that the nuclear X-ray source has a luminosity
L$_x$=$8\times10^{40}$ erg s$^{-1}$ in the 0.3--8~keV band and a slope
$\Gamma$=1.88$\pm$0.09 (Fabbiano et al. 2003).

In carrying out the spectral analysis we have accounted for the presence of
the AGN in IC~1459 by introducing a variable slope power law component with
a variable absorption in our spectral fits for the central region.
The presence of the bright point-source at the center of the group
complicates the spectral analysis. We find a significant
temperature variation $3^\prime$ distance from the center, rising from
$0.3\pm0.1$ to $1.0\pm0.2$ keV, with the hotter 
temperature observed to the north-west and south-east of the
galaxy. Additional attempts to further quantify the substructure in this
compact system were stymied by the point source in the center, the
relatively large wings in the PSF, and the low number of counts in 
the diffuse gas. Outside the region of the temperature rise, the group gas
has a constant temperature profile, at least within the temperature error,
while the other fitted quantities can vary significantly. But with the
contribution of the scattered photons from the central point source we
reserve judgement on the physical nature of the variations. In addition
to the XMM data there is also a Chandra observation of this system and 
it may provide insight into the diffuse gas but that analysis is beyond 
the scope of this paper.

\subsection{NGC~3923, $z=0.0069$}

The central galaxy of this group, NGC~3923, is a nearby early-type galaxy 
classified as E4 (de Vaucouleurs et al. 1991). This galaxy has numerous
shell structures (Malin \& Carter 1983) indicating past merger activity
but shows no signs of an AGN. 

The X-ray surface brightness of this group has a regular appearance and the
X-ray emission can be traced to about $5^{\prime}\,$. This group shows the
presence of numerous point sources which somewhat complicates the analysis.
NGC~3923 has previous {\sl ROSAT} PSPC and HRI data. The PSPC spectral
results are for two zones out to 2\arcmin$\,$ and show no evidence for a
temperature gradient (Buote \& Canizares 1998) at the level of $0.50\pm0.05$
keV. The {\sl ASCA} data for a slightly larger region indicate a slightly
higher best fit temperature of $0.64\pm0.07$ keV (Sato \& Tawara 1999), yet
a two-component fit yields a temperature of 0.55 keV (Buote \& Fabian
1998). The XMM data reveals a temperature structure within 2\arcmin$\,$
consisting of patches of lower ($0.35\pm0.05$ keV) temperature and an
approximately constant temperature outside this region at the level of
$0.56\pm0.01$ keV.  We note that we include a hard component as part of the
fit to account for a contribution of unresolved LMXB.

The results of the spectroscopic analysis reveal that variations in the
temperature are associated with the presence of the gas with entropy lower
than the value expected by interpolating other points.  Such low-entropy
regions could, for example, be produced by gas entrainment during preceding
jet activity and have not yet settled down.  An alternative explanation may
be that merger activity can inject low entropy gas. The statistical
analysis, reported in Tab.\ref{t:scatter} concludes that the statistically
significant fluctuations are only seen in the entropy.

Overall, the pressure map shows a fairly symmetric structure. There is a
small enhancement 0.5 arcminute to the north-east of the galaxy, which is
blended with the central emission and cannot be separated spectrally for
an independent analysis. This enhancement is visible on the temperature map
and may show up faintly in the entropy map. Overall the entropy map appears
quite symmetric as well.

The overlay of the X-ray surface brightness contours on the optical image
shows the enhancement in the pressure corresponds to the direction of the
major axis of the galaxy and therefore might be associated with triaxiality
of the gravitational profile.

\subsection{NGC~4168, $z=0.0077$}

The two peaks in the pressure and entropy maps correspond to the two bright
galaxies in the XMM field of view. The pressure map does not show any sign
of hydrodynamic interaction between those two galaxies and is elongated to
the north, possibly indicating the direction of the gas accretion.

Lack of counts precludes a detailed analysis of this group. The galaxy to
the west (NGC~4164) turns out to be a background galaxy ($V_H=17500$, while
the central galaxy in this group, NGC~4168, has a velocity of $V_H=2784$).
While a discrimination of the background galaxy is trivially done by
separating them in the velocity space, the important point is that we were
able to separate the effects of projection from the interaction.

\subsection{NGC~4261, $z=0.0068$}

X-ray emission of NGC~4261, { also reported in Gliozzi et al. 2003,}
extends over the whole XMM field of view.  Both pressure and entropy maps
appear symmetric with some elongation in the north-south direction.

\includegraphics[width=8.cm]{f6.ps}
\figcaption{The surface brightness profile in NGC4261. The circles (stars)
demarcate the profile in the north-south (east-west) direction. The figure
zooms in to the spatial region, where differences are seen, while the
profiles are identical within the central $20^{\prime\prime}$ and outside
$3^\prime$.
\label{n4261:pro}}

\includegraphics[width=8.cm]{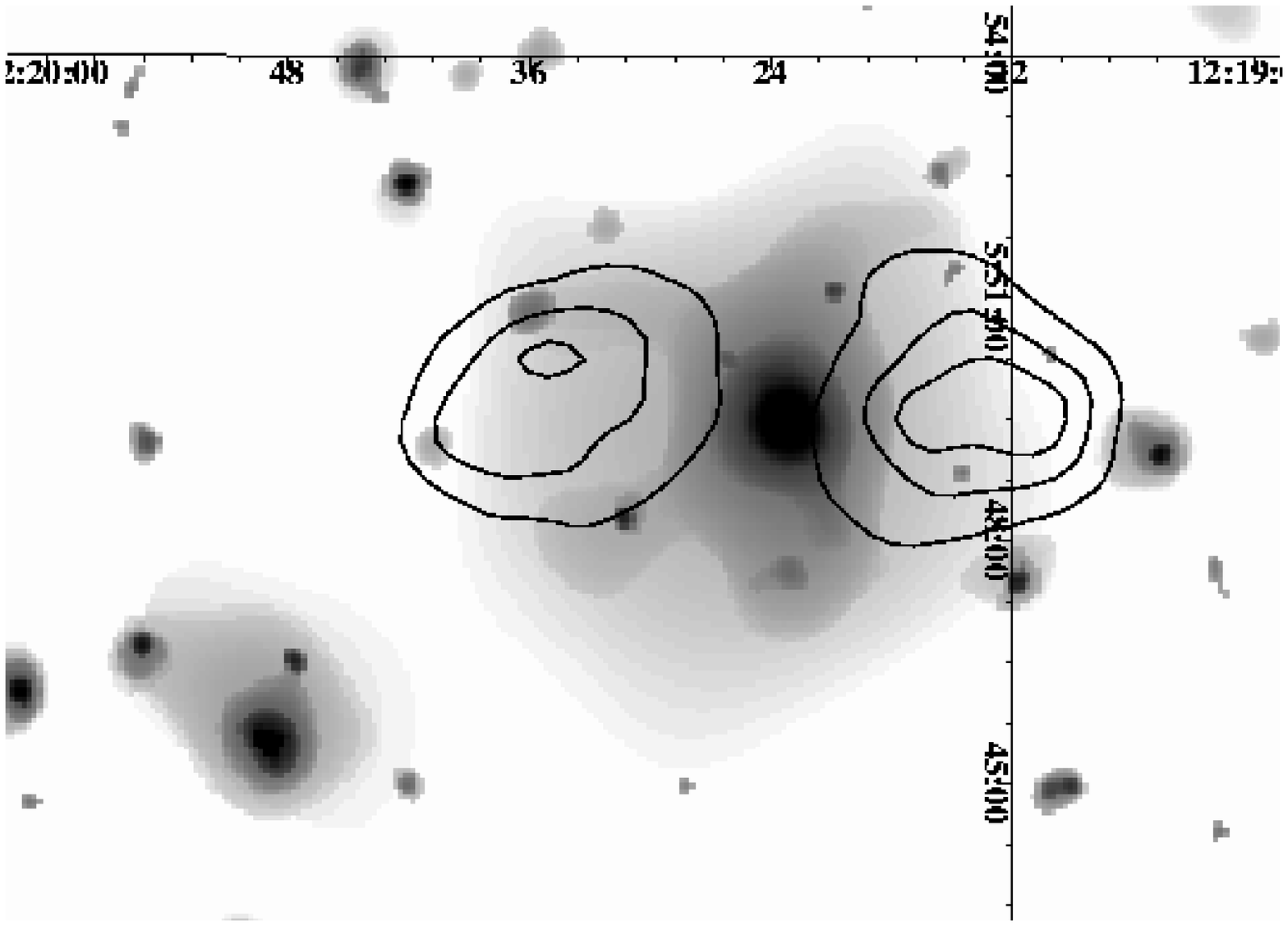}
\figcaption{Wavelet reconstructed X-ray image of NGC~4261, overlaid
with the radio contours.
\label{n4261:radio}}

The spectroscopic analysis confirms the extension in the pressure in the
north and south directions.  Overall, the very high temperature (1.--1.5 keV)
and low luminosity of the system yield very high entropy levels.

The differences between the north-south and the east-west directions are
clearly seen in the surface brightness reported in Fig.\ref{cf:1} as well as
in the profile shown in Fig.\ref{n4261:pro}. The profiles are identical
within the central $20^{\prime\prime}$ and outside $3^\prime$.  An
explanation to the peculiar emission in NGC~4261 is offered through the
comparison with the radio data, as is seen in Fig.\ref{n4261:radio}, the
radio lobes fill the space between the X-ray emission. The presence of the
radio bubble introduces an additional pressure at the level of
$2\times10^{-12}$ dyne cm$^{-2}$. Detailed analysis shows that the thermal
gas inside the eastern bubble is in pressure equilibrium with its immediate
surroundings. The gas inside the bubbles is characterized by a higher
entropy $98\pm9$ keV cm$^2$ to the east and $105\pm11$ keV cm$^2$ to the
west compared with $85\pm6$ keV cm$^2$ at similar distance to the center as
well as $30\pm3$ keV cm$^2$ in the central 10 kpc. The western bubble has
lower pressure compared to surroundings, indicating an importance of the
relativistic particle energy.  Our measurements support the claim of Croston
et al. (2005) of finding the direct evidence of the AGN reheating of gas in
groups. We note a remarkable similarity in the optical properties of NGC
4261 and NGC 5322, they have similar absolute B magnitudes (derived using
the distance measurements from Tonry et al. 2001 and the corrected B
magnitude from the RC3), -19.52 and -19.84
respectively, they are both classified as liners, and the effective radii
are similar, log r$_e$=1.59 and 1.54 respectively.  In addition the colors
are similar $\Delta (B-R)$ =0.03 as are the color gradients (Michard
1999). Yet, the entropy of the gas at the center of NGC~4261 is a factor of
four larger (the entropy in the core of NGC~5322 is $6.5\pm2$ keV
cm$^2$). As the gas properties at the center are dominated by the stellar
mass loss which is expected to be the same between the two galaxies, the
observed difference measures the degree of impact the AGN has on the
interstellar medium.

\subsection{NGC~4636, $z=0.0044$}

NGC~4636 has been classified as a very symmetric, E0, elliptical galaxy
(RC3) based on the optical analysis of the central region. However, at
larger radii the optical isophotes become increasingly elliptical (King
1978). The Einstein IPC image of this galaxy reflects the optical morphology
with very symmetric emission near the center and asymmetry on larger scales
(Forman et al. 1985; Stanger \& Warwick 1986).  The {\sl ROSAT} PSPC data
show a flattening of the X-ray surface brightness profile starting about
4$\arcmin\,$ to the north-east of the galaxy center (Trinchieri et al. 1994).

The XMM data show this flattening of the profile at the same position as 
the PSPC data. Examining the pressure map near this location 
we identify a peculiar inward-pointed bow-shape enhancement also $\sim$4
arcminutes to the north-east from the center. We find no associated
entropy dip with any of these pressure enhancements. This might indicate the
presence of a shock. Similar conclusions have been reported using Chandra
observations (Jones et al. 2002).

The entropy map shows complex structures which are well correlated with
structures seen in the pressure map in the inner 3$\arcmin\,$. At larger
radii the entropy map shows fewer distinct features.  One possible
explanation for the small scale features might be that foreground or
background objects might be contributing to the X-ray emission at these
locations. However, the DSS optical image does not reveal any optical object
corresponding to the identified pressure and entropy features.

\begin{figure*}

\mbox{\includegraphics[height=6cm,angle=-90]{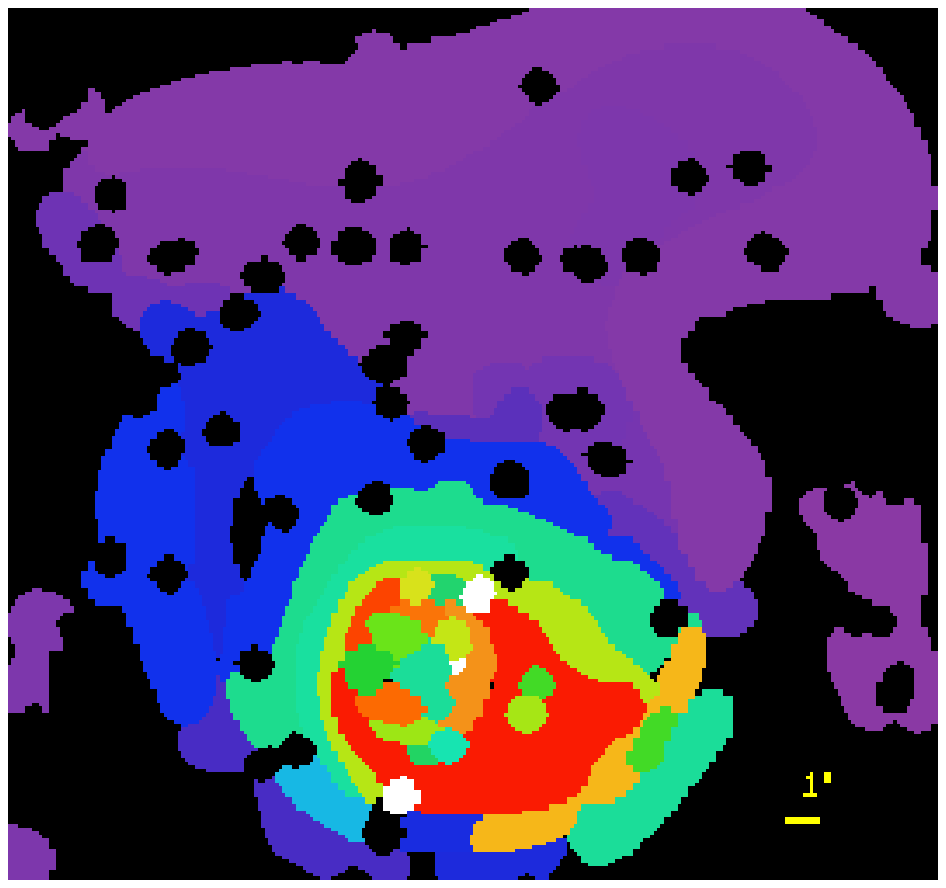}
\includegraphics[height=6cm,angle=-90]{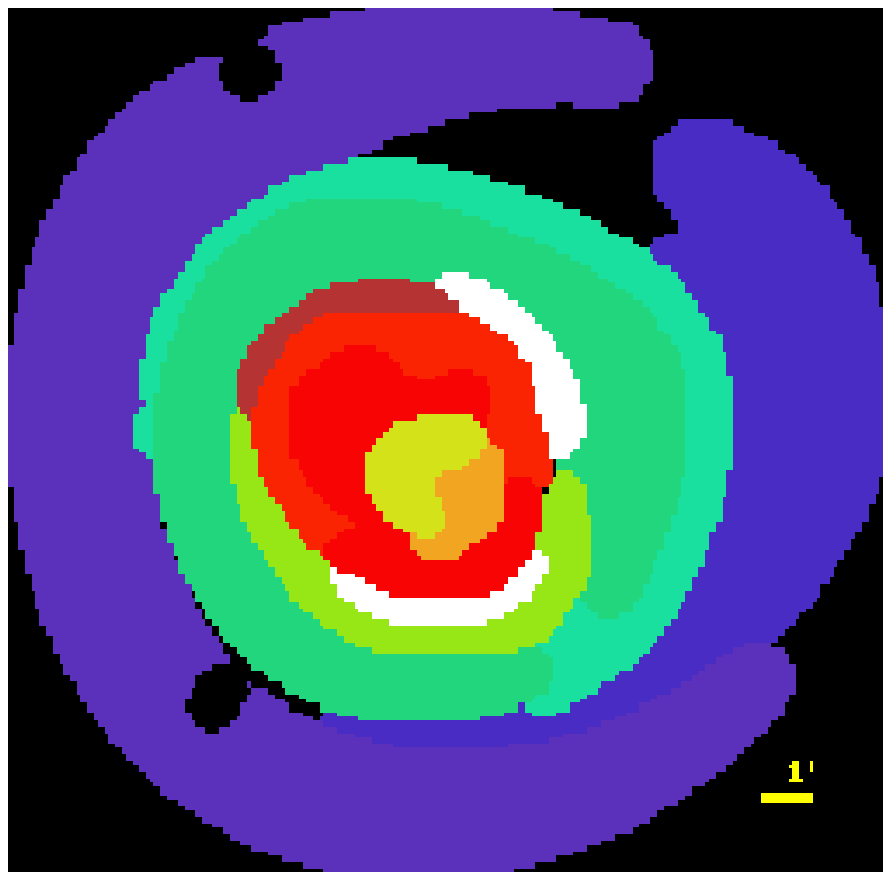}
\includegraphics[height=6cm,angle=-90]{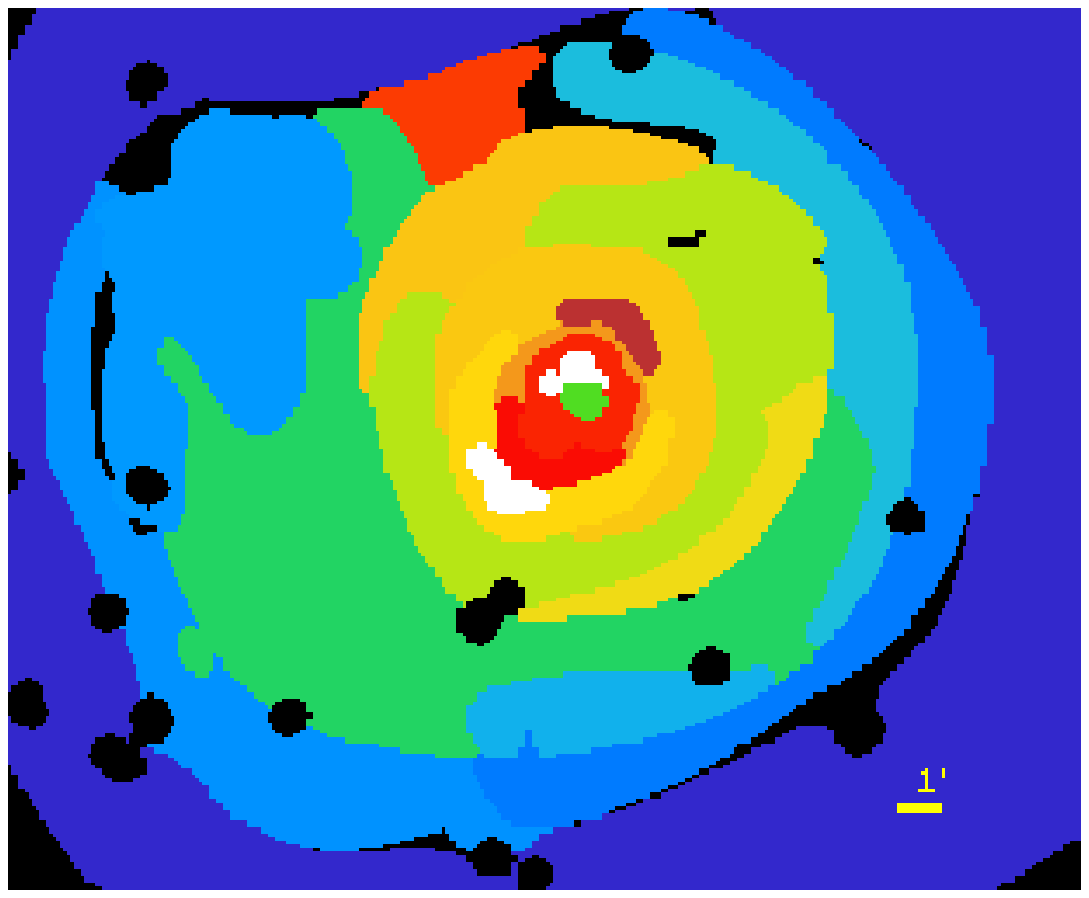}}

\figcaption{ Iron abundance maps of NGC~4636, NGC~5846 and NGC~5044 (from
the left to the right). Violet color represents the Fe abundance level of
0.1 solar, blue -- 0.2 solar, green -- 0.3 solar, red -- 0.5 solar and white
-- above 0.8 solar. A typical uncertainty is less than 0.1, which is smaller
than the color separation.
\label{f:femaps}}

\includegraphics[width=16cm]{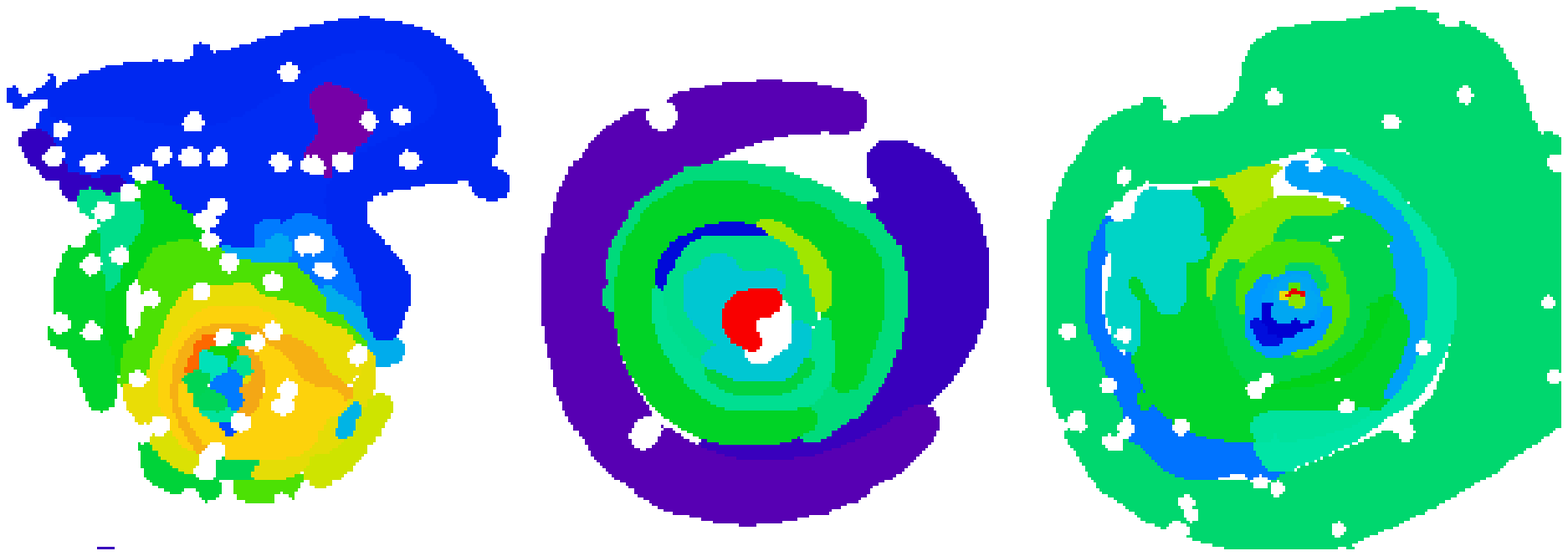}\\
\includegraphics[width=16cm]{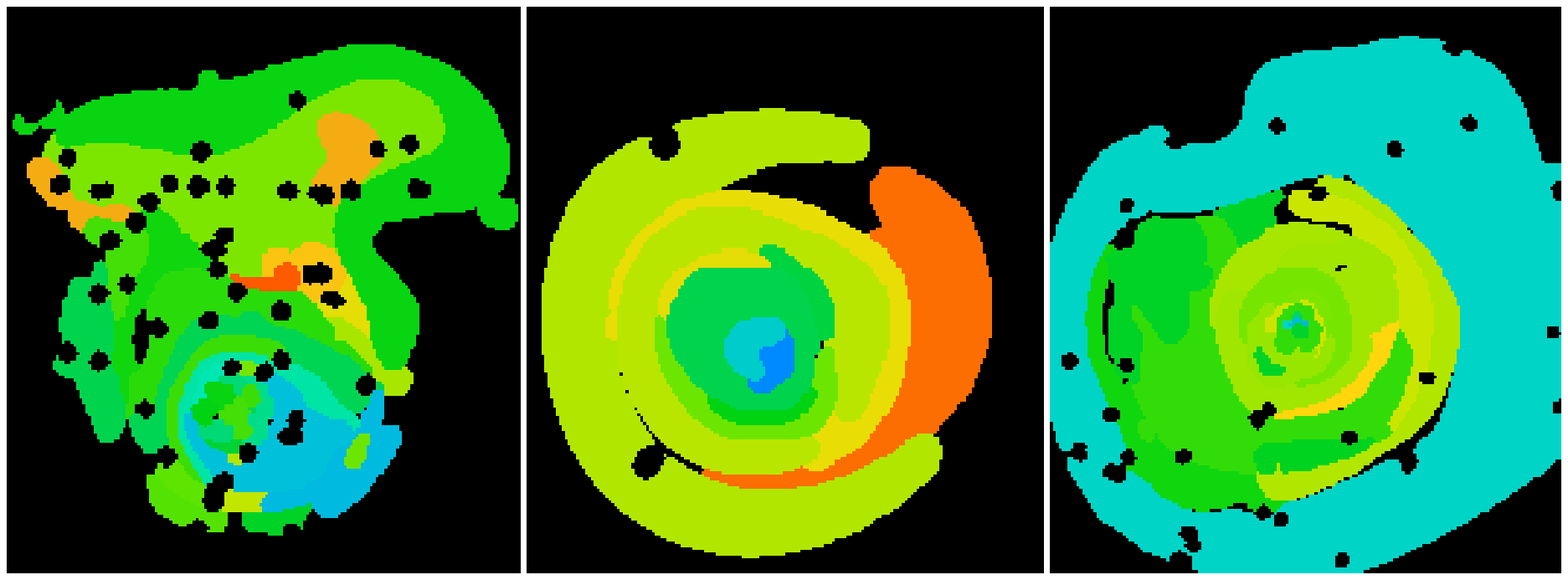}

\figcaption{The maps of ratio between the entropy (top), pressure (bottom)
and the mean trend of the corresponding quantity with radius, measured for
NGC~4636, NGC~5846, NGC~5044 (from the left to the right).
\label{f:sprats}}

\end{figure*}

The most striking large-scale feature of NGC~4636 consists of the strong
enhancement of the emission 10 arcminutes to the north from the center,
which we will refer to as a plume (Fig.\ref{f:sprats}). Although appearing
on the edge of the XMM-Newton field of view, this emission is also seen in
ROSAT data. The optical data do show that there is a small galaxy
concentration to the north of the galaxy. Redshift information from NED
shows that this group has a redshift of $z=0.07$, but this system does not
coincide with features in the maps. Also, an association with the polar spur
appears unlikely, as no peculiarity is seen in the IR maps.  Furthermore,
the parameters of the emission (0.6 keV) are not typical of that of the spur
(0.3 keV). In addition, we have examined the RASS data and this is not part
of a larger scale feature that extends beyond the XMM FOV.

The results of spectral analysis reported in
Fig.\ref{f:femaps}-\ref{f:sprats} show that the plume has a similar pressure
to the surroundings, and it has a low element abundance consistent with its
surroundings. However, the entropy is low when compared to the surrounding
gas and indeed shows some of the lowest entropy in the NGC~4636 system. One
explanation is that this is low-entropy group gas. Analysis of the pressure
map indicates an azimuthal variation in the pressure at $3^\prime$ radius
from the center, possibly related to the origins of the plume. Low entropy
gas is seen to the south and west of the galaxy center. The structure of
this low entropy gas is reflected in the iron abundance maps and can be seen
as a region of high abundance. The plume to the north of the galaxy is a
sign that NGC~4636 is undergoing stripping. Recently Chandra observation
have revealed a cold front on the southern part of NGC4636 (Jones, C. 2004,
private communication), which on one hand is an additional argument
supporting our association of the plume with the ongoing stripping process
and on the other, favors a core oscillation scenario (.e.g. Tittley \&
Henriksen 2005) as an origin of interaction. In the case of NGC4636, the
core oscillation could be caused by the gravitational pull of the Virgo
supercluster, a suggestion that could be tested using numerical
simulations. The iron abundance map, present in Fig.\ref{f:femaps}, reveals
a strong contrast in the metalicity. The metalicity inside the central
$1.5^\prime$ (5.9 kpc) varies between 0.5 and 1.0 with a typical error of
0.05 and a typical element resolution of 2 kpc. At $2.6^\prime$ distance
from the center the metalicity reaches $1.1\pm0.1$ value (shown in red). The
high-metalicity zone has a stronger extension towards the south-west. The
metalicity of the plume is much lower, $0.13\pm0.01$, but comparable to the
zone outside the core at a similar ($12^\prime$) distance to the center,
studied with annuli.

Our reinvestigation of the group membership suggest absence of galaxies at
velocities similar to NGC~4636. On the other hand, the arguments of the
X-ray analysis infer that NGC~4636 has a group-size halo, which confirms a
conclusion of Matsushita et al. (1998) attributing NGC~4636 to fossil
groups, a recently emerged class of objects (Ponman \& Bertram 1993;
Vikhlinin et al. 1999; Mulchaey \& Zabludoff 1999).

\subsection{NGC~5044, $z=0.0082$}

The two dimensional spectral mapping of the properties of the X-ray emission
from NGC~5044 is done at the effective 4--10 kpc resolution, within the
central 50 kpc. The temperature map of NGC~5044 (Fig.\ref{cf:1}) reveals a
cool zone one arcminute south-east from the X-ray center. The parameters of
this zone are the following: $M_{\rm gas}=2\times10^9 M_\odot$,
kT$=0.82\pm0.01$ keV, Fe$=0.9\pm0.1$Fe$_\odot$, S=$19\pm1$ keV cm$^2$,
P$=1.1\pm0.2\times10^{-11}$ dyne cm$^{-2}$. This feature has been seen in
the analysis of the ROSAT data by David et al. (1994), who associate its
appearance with a cooling wake. The idea was that motion of NGC~5044 in the
cooling gas of its group creates an enhancement of the dropping-in gas
behind the galaxy. However, an assumption of the cooling flow, used in that
scenario is no longer a prefered interpretation of the data (e.g. Tamura et
al. 2001). XMM-Newton data, shown in Figs.\ref{f:femaps}--\ref{f:sprats}
reveal that the wake has an entropy lower compared to both the mean trend
for the atmosphere of NGC~5044 as well as the entropy of the gas located on
the opposite side from the center. On the other hand, the entropy of the
wake is similar to the entropy of the gas at the center, which suggests a
similar origin. The characteristics of the central region where a similar
mass of the gas is encompassed are the following: kT$=0.81\pm0.01$ keV,
Fe$=0.82\pm0.05$Fe$_\odot$, S=$14\pm1$ keV cm$^2$,
P$=1.9\pm0.1\times10^{-11}$ dyne cm$^{-2}$. The parameters of the gas at the
same distance to the center as the wake are: kT$=0.97\pm0.01$ keV,
Fe$=0.66\pm0.05$Fe$_\odot$, S=$29\pm1$ keV cm$^2$,
P$=0.97\pm0.05\times10^{-11}$ dyne cm$^{-2}$. High metalicity of the wake,
detected by XMM-Newton, also supports a conclusion of the same origin of the
gas in the wake and in the center of NGC~5044. After reexamining the origin
of the wake, we conclude that both the motion of the galaxy (e.g. due to the
core oscillations) and former AGN activity can provide an explanation. The
extent and form of the entropy wake of NGC~5044, shown in Fig.\ref{f:sprats}
supports the gas stripping hypothesis. The reported deviations bear a local
character, while the statistical analysis run within the $5^\prime$ radius
from the center (Tab.\ref{t:fe}) confirms the regularity in the pressure
distribution. A dispersion in the Fe abundance (Tab.\ref{t:scatter}) is
higher than the dispersion in the entropy (Tab.\ref{t:scatter}), indicating
different origins of the gas that shares the same state. Indeed, stellar
mass loss of the central galaxy, stripping of the group members as well as
low-entropy part of the intragroup medium are very close in their gas
characteristics, yet poses a different metalicity. We conclude that
XMM-Newton supports the work on using the metalicity as a tracer of the
formation of IGM (e.g. Schindler et al. 2004).

\subsection{NGC~5322, $z=0.0065$}

The central galaxy NGC~5322 is a bright, E3 galaxy (de Vaucouleurs et
al. 1991). Early radio observations show that this galaxy is a weak radio
source (Feretti \& Giovannini 1980; Hummel 1980) with symmetric jets
emanating from an unresolved core located at the optical center (Feretti et
al. 1984).

The bright part of the X-ray emission is rather compact ($1^\prime$), peaked
on NGC~5322, and extends only to the south of the galaxy. There is a dip in
the entropy map $\sim$ 1.2 arcminutes to the south-west from the center with
no optical counterpart in DSS2 images. { Insufficient quality of the data
only allowed us to select 6 regions for the spectral analysis, precluding us
from confirming the entropy dip. The enhancement in the pressure
$3.7^\prime$ to the south-east of NGC~5322 is only marginally seen in the
spectral analysis.}

\subsection{NGC~5846, $z=0.0063$}

This system reveals a number of ring-like structures in the image, which we
identify with metalicity enhancements. No optical counterpart for this
structure is seen in the DSS2 images, consistent with the idea of metalicity
enhancements. Although an optical image reveals a small galaxy
$0.6^\prime$ south to NGC~5846 (NGC~5846a, velocity difference 500 km/s), it
is not demarcated on the entropy map. This rules out a significant gas halo
for this object and so it should have in any case a negligible impact on the
X-ray appearance.

The two dimensional spectral mapping of the properties of the X-ray emission
from NGC~5846 is done at the effective 6--10 kpc resolution, within the
central 50 kpc. As the spectroscopic analysis shows (see
Fig.\ref{f:sprats}), entropy, pressure, temperature and iron abundance all
appear relaxed with only a slight displacement between the central peak and
the center of the large-scale emission. The large-scale elongation of the
entropy is confirmed and an associated elongation in the high-element
abundance zone is found. In addition, some disturbance in the pressure map
is observed in the core.  Given the long time scale for the entropy
disturbance to relax, it is not surprising that the associated radio
activity is not observed. The presence of an AGN in NGC~5846 is supported by
the detection of a central radio point source in the NRAO VLA Sky Survey
(Becker et al. 1995). { There is a zone of high Fe abundance $2^\prime$ to
the north-east from the center (shown in red in Fig.\ref{f:femaps}, on the
level of $0.75\pm0.07$ solar, compared to $0.50\pm0.05$ detected at a
similar distance from the center, which is a base of our claim on
metalicity variations. Patches of higher abundance, shown in white in
Fig.\ref{f:femaps} correspond to a $1.0\pm0.3$ values and are of marginal
significance.}

Another possibility is that the structures seen in NGC~5846 are sound waves,
as have recently been reported for the Perseus cluster (Fabian et al. 2003).

\subsection{NGC~7582, $z=0.0054$}

The $\sim5^\prime$ extent of the X-ray emission of NGC~7582 is shown in
Fig.\ref{f:ol}. Two other galaxies, NGC 7590 and NGC 7599, (not shown in
Fig.\ref{f:ol}) also reveal extended X-ray emission on the scales of
$1-2^\prime$, each one centered on a galaxy. Our study confirms the lack of
diffuse X-ray emission from this group reported by Mulchaey \& Zabludoff
(1998) based on the ROSAT data. NGC~7582 is a well studied Seyfert 2
galaxy. Its hard (2-10 keV) X-ray emission has been detected in many X-ray
surveys (Ward et al. 1978) and the X-ray spectrum of this AGN is complex
(c.f. Turner et al. 2000). Detailed spectroscopic analysis shows a presence
of emission from the AGN and complex diffuse emission from the bright spiral
galaxy (NGC~7582). Since the origin of the emission from the spiral galaxy
is likely to be a mixture of emission from hot gas, point sources and
supernova remnants, which is quite different from the rest of the sample, we
omit this system from further analysis.

In Koribalski (1996) the HI image of the NGC~7582 group was presented. A
comparison shows that the X-ray emission is embedded within the $H_\alpha$,
and that the shape of the X-ray emission is determined by the star-formation
processes, as opposed to the case where the hot gas fills the potential of
the galaxy.

\section{Summary and Conclusions}

We performed an innovative 2-dimensional analysis of the IGM properties in a
representative sample of low-z groups of galaxies. We have well sampled the
transition region between the regular groups and the X-ray faint groups and
also have two examples of a strong AGN feedback.

The properties of the hot gas in the X-ray faint groups are studied at radii
reaching $0.3r_{500}$ for the first time. The properties of the hot gas in
NGC~5322 and NGC~3923 deviate from the relations derived using brighter
groups and clusters even at largest radii. These faint systems having higher
entropy and lower pressure than expected.  Our findings suggest that the
X-ray emitting gas in roughtly half of X-ray faint groups (groups with
bolometric $L_x<10^{41}$ ergs s$^{-1}$) does not originate from infalling
gas heated to X-ray temperatures but instead is a byproduct of the
galaxy-galaxy merger activity associated with the formation of an elliptical
galaxy. In fact, we note a similarity in the properties of the X-ray faint
groups studied here and the X-ray appearance of the post-merger galaxies.

Four other X-ray faint groups, IC~1459, NGC~2300, NGC~4168 and NGC~4261,
exhibit typical group IGM properties outside $0.2 r_{500}$ and we consider
them as regular groups. One such group, NGC4261, also reveals AGN activity
and our study reveals faint X-ray cavities, filled with the relativistic
plasma, extending to a similar distance from the center, which are also
reported in Croston et al. (2005) and confirms their conclusion that AGN
activity has a substantial impact on the state of the gas at the central
part of groups.

Examination of the structure in the gas of three groups reveals fluctuations
in the entropy and pressure on a typically 10-30\% and metalicity on the
30--50\% level. The larger scatter in the metalicity of the gas within the
group, compared to its entropy suggests a diversity in the origin of the gas
in the core and supports the work on using the metalicity as a tracer of the
formation of IGM. However, use of metalicity measurements is currently
limited to a small number of systems, due to requirement of a large number
of detected counts in each region. In our study, such analysis is only
possible for NGC4636, NGC5044 and NGC5846.

The two dimensional information, related to the analysis reported in this
paper, is released under http://www.mpe.mpg.de/2dXGS/ homepage.

\section{Acknowledgments}
The paper is based on observations obtained with XMM-Newton, an ESA science
mission with instruments and contributions directly funded by ESA Member
States and the USA (NASA). The XMM-Newton project is supported by the
Bundesministerium f\"{u}r Bildung und Forschung/Deutsches Zentrum f\"{u}r
Luft- und Raumfahrt (BMFT/DLR), the Max-Planck Society and the
Heidenhain-Stiftung, and also by PPARC, CEA, CNES, and ASI. The authors
thank an anonymous referee for an insightful referee report, which led to
substantial improvements in the quality of presented material.  AF thanks
Joe Mohr, Stefano Borgani, Trevor Ponman, and Mark Henriksen for useful
discussions. AF thanks Steve Helsdon for his help in the regression
analysis. AF acknowledges support from BMBF/DLR under grant 50 OR 0207, MPG
and a partial support from NASA grant NNG04GF686. AF and MZ thank the Joint
Astrophysical Center of the UMBC for the hospitality during their visit. DSD
acknowledges partial support for this project from NASA grant NAG5-12739. JSM
acknowledges partial support for this program from NASA grant NNG04GC846.

\end{document}